\documentclass[prl,aps,twocolumn,superscriptaddress]{revtex4-1}
\makeatletter
\usepackage{pdfpages} 
\AtBeginDocument{\let\LS@rot\@undefined}
\makeatother

\usepackage{amsmath,amssymb}
\usepackage{graphicx}
\usepackage{epsfig, color}
\usepackage{wasysym}
\usepackage{amsfonts}
\usepackage{bm}
\usepackage{enumerate}
\usepackage{bbold}
\usepackage{xcolor}
\usepackage[normalem]{ulem}
\usepackage{cancel}
\usepackage{multibib}
\usepackage{comment}

\usepackage{mathtools}

\usepackage{latexsym}
\usepackage[breaklinks,colorlinks = true,linkcolor = red,urlcolor=cyan,citecolor=red]{hyperref}

\newcommand{\be}{\begin{eqnarray}}
\newcommand{\ee}{\end{eqnarray}}

\newcommand{\ident}[0]{ \mathbb{1}}
\newcommand{\swap}[0]{ \mathbb{P}}

\DeclareMathAlphabet{\mathpzc}{OT1}{pzc}{m}{it}

\def\evol{\hat{M}}
\def\Floq{\hat{W}}
\def\Tmat{\hat{\mathbb{T}}}
\def\Rmat{\hat{\mathcal{R}}}
\def\Ham{\hat{\mathbb{H}}}
\def\Transfer{\hat{\mathcal{T}}}
\def\gate{\hat{U}}
\def\jp{j^{\prime}}
\def\disp{\varepsilon}

\newcommand{\vpd}[0]{\vphantom{\dagger}}
\newcommand{\vps}[0]{\vphantom{*}}
\newcommand{\tth}[0]{t^{\vpd}_{\rm Th}}
\newcommand{\theis}[0]{t^{\vpd}_{\rm Heis}}
\newcommand{\Hdim}[0]{\mathcal{D}}

\newcommand{\expval}[1]{ \left\langle #1 \right\rangle}

\newcommand{\Trace}{{\rm Tr}}

\newcommand{\abs}[1]{\left| #1 \right|}

\def\ket#1{{|#1\rangle}}

\def\HXXX{H_{\rm XXX}}

\usepackage{graphicx}

\newcommand{\titleinfo}{Spectral statistics and many-body quantum chaos with conserved charge}

\begin{document}
\title{\titleinfo}
\author{Aaron J. Friedman}
\affiliation{Rudolf Peierls Centre for Theoretical Physics, Clarendon Laboratory, University of Oxford, Oxford, OX1 3PU, UK}
\affiliation{Department of Physics and Astronomy, University of California, Irvine, CA 92697, USA}
\author{Amos Chan}
\affiliation{Rudolf Peierls Centre for Theoretical Physics, Clarendon Laboratory, University of Oxford, Oxford, OX1 3PU, UK}
\author{Andrea De Luca}
\affiliation{Rudolf Peierls Centre for Theoretical Physics, Clarendon Laboratory, University of Oxford, Oxford, OX1 3PU, UK}
\affiliation{Laboratoire de Physique Th\'eorique et Mod\'elisation (CNRS UMR 8089), Universit\'e de Cergy-Pontoise, F-95302 Cergy-Pontoise, France}
\author{J. T. Chalker}
\affiliation{Rudolf Peierls Centre for Theoretical Physics, Clarendon Laboratory, University of Oxford, Oxford, OX1 3PU, UK}
\date{\today}

\begin{abstract}
We investigate spectral statistics in spatially extended, chaotic many-body quantum systems with a conserved charge. 
We compute the spectral form factor $K(t)$ analytically for a minimal Floquet circuit model that has
a $U(1)$ symmetry encoded via 
spin-$1/2$ degrees of freedom.
Averaging over an ensemble of realizations, we relate $K(t)$ to a partition function for the spins, given by a Trotterization of the spin-$1/2$ Heisenberg ferromagnet. 
Using Bethe Ansatz techniques, we extract the `Thouless time' $ \tth $ demarcating the extent of random matrix behavior, and find scaling behavior governed by diffusion for $K(t)$ at $t\lesssim \tth$. We also report numerical results for $K(t)$ in a generic Floquet spin model, which are consistent with these analytic predictions. 
\end{abstract}

\maketitle

\noindent{\textit{Introduction.}} 
Statistical mechanics is a fundamental tool in understanding condensed matter systems, allowing for their description in terms of a few state variables, rather than thermodynamically many degrees of freedom. For quantum or classical systems in equilibrium with their environment, thermodynamics arises naturally from the exchange of conserved quantities with a thermal reservoir. For generic isolated systems, thermalization is not guaranteed, but rather must emerge dynamically. 
In recent years, there has been substantial theoretical~\cite{Rigol} and experimental effort~\cite{RevModPhys.80.885, bloch2012quantum} to understand how many-body quantum systems in \emph{isolation} equilibrate under their own dynamics to reproduce the familiar results of statistical mechanics. 
The Eigenstate Thermalization Hypothesis (ETH) \cite{ETH1,ETH2,Rigol2008kq}  
provides a universal mechanism for establishing ergodicity of isolated quantum systems: in its simplest formulation, it is based on the convergence of the expectation values of local observables in nearby energy eigenstates when the thermodynamic limit is considered. 
Indeed, this assumption is enough to recover thermal behavior from the long-time dynamics of many-body quantum systems. 
The notion of quantum ergodicity associated with ETH is intertwined with \textit{random matrix theory} (RMT): first, quantum chaotic systems 
are characterized by an RMT eigenvalue distribution~\cite{bohigas1984characterization, guhr1998random}; second, their eigenfunctions can be understood as random vectors~\cite{berry1977mv, ETH1,ETH2, borgonovi2016quantum}.
One consequence is that quantum thermalization is always associated with level repulsion between energy eigenvalues. In practice, this \textit{spectral rigidity} has often been used as an efficient means to pinpoint quantum ergodicity breaking~\cite{rigol2009breakdown, santos2010onset, biroli2010effect, pal2010many, alba2015eigenstate}.

The validity and possible regimes of violation of ETH have been scrutinized in different types of chaotic systems~\cite{rigol2012alternatives, luitz2016anomalous, dymarsky, foini2019eigenstate, foini2019rotational, CDLC3}; however, numerical tests of ETH are challenging as they require the diagonalization of Hamiltonians whose size grows exponentially with the number of microscopic degrees of freedom~\cite{brandino2011quench, MoessnerHaque, kim2014testing, luitz2016anomalous}. An obvious limitation is that, while RMT captures several aspects of quantum chaos, replacing the microscopic time evolution by a random matrix overlooks a key facet of the former, namely locality. 

Recent efforts have endeavored to establish and improve upon minimal models of chaotic many body quantum systems starting from RMT, and enforcing locality via random \emph{local} unitary gates to form `circuits' \cite{NahumRUC1,RUCNCTibor,CDLC1,CDLC2,CDLC3,RUCconTibor,RUCconVedika,CiracRUFC,ProsenRMTChaosPRX,ShenkerRMT}. 
Such models generally display quantum chaos, as characterized by entanglement entropy, the decay of local observables, and out-of-time-ordered correlation functions \cite{NahumRUC1,RUCNCTibor,CDLC1}.
By considering Floquet random circuits, it has been possible to derive analytically RMT spectral rigidity, in the limit of large local Hilbert space dimension~\cite{CDLC1, CDLC2} or at fine-tuned solvable points~\cite{ProsenRMTChaosPRX, bertini2018exact}. More precisely, these works demonstrated that RMT behavior only appears for eigenvalue separations 
small on the scale of the inverse of
the Thouless time, $\tth$, named in analogy with single-particle disordered conductors~\cite{ThoulessPRL,AltshulerShklovskii}.
The value of $\tth$ depends on the linear system size $L$ and characterizes the time scale for the onset of quantum chaos. It remains an open question to understand which mechanisms control the scaling of $\tth$ with $L$.  

In this work, we investigate the effect of a local conserved quantity $\hat Q$ on the behavior of $\tth$ and, more generally, on the spectral properties of the evolution operator, $\Floq$, of a Floquet system. In particular, we consider the two-point spectral form factor (SFF) $K(t)$, defined as
\be \label{Ktdef}
K(t) \equiv \sum\limits_{m,n = 1}^\Hdim \expval{e^{\imath \left( \theta_m - \theta_n \right) t }} = \langle |\Trace[\hat W(t)]|^2\rangle\,.
\ee
Here $\{\theta_m\}$ are the eigenphases of $\Floq$, $\Hdim$ is the Hilbert space dimension, $\hat W(t)$ indicates the $t$-th power of $\hat W$, and 
$\langle \ldots \rangle$ denotes the average over an ensemble of statistically similar systems. The SFF is the Fourier transform of the two-point correlation function of eigenphases. For uncorrelated eigenphases, $K(t) = \Hdim$, while for random matrices belonging to the Circular Unitary Ensemble (CUE), $K_{\rm CUE}(t) = \abs{t}$ until the Heisenberg time $\theis = \Hdim$, after which $K_{\rm CUE}(t) =\Hdim$. The linear ramp is thus a fingerprint of level repulsion. For spatially-extended one-dimensional (1$d$) systems without a conserved density~\cite{CDLC2,GuhrArxiv}, $K(t) \simeq t^{L/\xi(t)}$ for $t \ll \tth$: the system can be seen as partitioned into $L/\xi(t)$ chaotic blocks, with a length $\xi(t)$ that grows with $t$. RMT behavior is recovered for $t \gtrsim \tth$ with $\xi(t = \tth) \sim L$. 
In the presence of a conserved quantity with diffusive transport, it is natural to expect 
$\tth \sim L^2 / D$, where $D$ is the diffusion constant. 
The idea that the timescale $L^2/D$ controls the onset of RMT spectral correlations was proposed on a heuristic basis in \cite{ShenkerRMT}, with support from a variety of estimates and numerical studies. Here we establish this result in an exact treatment of a minimal model. We also set out the scaling behavior of $K(t)$ in the time interval $1\ll t \lesssim \tth$ and show that this holds in a computational study.

To probe $K(t)$ we build on a Floquet circuit model introduced in \cite{CDLC1}, consisting of a chain with $q$-state `spins' at each site. The model has a time-evolution operator $\Floq$ constructed from unitary gates that act on neighboring pairs of sites. Gates are randomly selected in space but repeated periodically in time. Using a diagrammatic method to average over the individual matrices \cite{BnB}, to leading order at large $q$, one finds $K(t) = K_{\rm CUE}(t)$ for any $t \neq 0$, so that $\tth \to 0$ as $q \to \infty$. 
In the following, we formulate and characterize an extension of this model that hosts a $U(1)$ symmetry corresponding to a local, conserved operator $\hat Q$ that commutes with $\Floq$. In this way, the limit $q \to \infty$ has a twofold convenience: first, it allows controlled diagrammatic calculations; second, it washes out any effect on $K(t)$ not due to $\hat Q$. 

\noindent{\textit{Circuit model.}} 
The minimal model is a Floquet random unitary circuit (FRUC) defined on a chain of $L$ sites with local Hilbert space $\mathcal{H}_{\rm loc} \equiv \mathbb{C}^q \otimes \mathbb{C}^{2}$ -- the tensor product of a $q$-dimensional \textit{color} and a spin-$1/2$. The former facilitates Haar averaging~\cite{CDLC1,CDLC2,RUCconVedika}, and we encode a $U(1)$ symmetry in the latter, following \cite{RUCconVedika}, corresponding (in standard notation) to conservation of $\hat Q = \hat{S}^z = \frac{1}{2} \sum_{j=1}^{L} \hat{\sigma}^{z}_j$.

The single-period -- or Floquet -- evolution operator $\Floq\equiv \Floq^{\vps}_2 \cdot \Floq^{\vps}_1$ is a depth-two circuit comprised of local two-site gates: assuming even $L$, the two layers correspond respectively to odd and even bonds, with $\Floq^{\vps}_1 = \gate^{\vps}_{1,2} \otimes \gate^{\vps}_{3,4} \otimes \dots $ and $\Floq^{\vps}_2 = \gate^{\vps}_{2,3} \otimes \gate^{\vps}_{4,5} \otimes \dots \gate^{\vps}_{L,1} $. We require that each $\gate^{\vps}_{j,j+1}$ preserves the local magnetization $S^z_{j,j+1} = \frac{1}{2}\left(\hat{\sigma}^z_j + \hat{\sigma}^z_{j+1}\right)$~\cite{RUCconTibor,RUCconVedika}. Thus $\gate^{\vps}_{j,j+1}$ is a $4 q^2 \times 4 q^2$ block diagonal matrix, acting as a $q^2 \times q^2$ matrix in each of the $\uparrow \uparrow$ and $\downarrow \downarrow$ subspaces, and as a $2  q^2 \times 2 q^2$ matrix in the $\uparrow \downarrow \, , \, \downarrow \uparrow$ subspace, with all three blocks independently drawn Haar random unitaries.

To characterize spectral correlations in this quantum circuit, we compute the SFF \eqref{Ktdef}. Since $[\Floq, \hat{S}^z] = 0$, $\Floq$ is block-diagonal,
and levels from different $\hat{S}^z$ sectors do not repel. Thus, we define 
\be \label{eq:Kdef} K(t, s) \equiv \expval{ \underset{s}{\Trace}[\hat{W} (t)] \, \underset{s}{\Trace}[ \hat{W}^{\dagger} (t) ] } \ee
where `$s$' indicates restriction to the subspace $\hat S^z=S = L \, s$~\footnote{For finite $L$, only a discrete set of values of $s$ are allowed.} and $\expval{\cdots}$ denotes Haar averaging.

\noindent{\textit{Effective spin-$1/2$ model.}---} The ensemble averaging in \eqref{eq:Kdef} maps $K(t,s)$ to the partition function of a Trotterized Heisenberg ferromagnet. Evaluating this average amounts to generating all diagrams~\cite{CDLC1} by pairing unitaries $\gate^{\vps}_{j,j+1}$ with their complex conjugates $\gate^{\dagger}_{j,j+1}$ at each bond~\footnote{See supplementary material at [url].}. 
As $q \to \infty$, the leading contributions come from $t$ diagrams, each of which has an identical `cyclical' pairing at all sites \cite{CDLC1}. 
These diagrams can be expressed algebraically as (see \cite{Note2} for details)  
\be \label{eq:Ktrace} \lim_{q\to \infty} K(t, s) = \abs{t} \underset{s}{\Trace}[ {\evol}^{\, t} \,] \, ,  \ee
where the factor of $\abs{t}$ comes from there being $t$ such leading diagrams. The trace over the effective spin-$1/2$ evolution operator, $\evol$, accounts for the sum over the color and spin degrees of freedom in a given leading diagram. Like $\Floq$, $\evol \equiv {\evol}^{\vps}_2 \cdot {\evol}^{\vps}_1 $ consists of two layers:~${\evol}^{\vps}_1 = {\Tmat}^{\vps}_{1,2} \otimes {\Tmat}^{\vps}_{3,4} \otimes \dots $ and ${\evol}^{\vps}_2 = {\Tmat}^{\vps}_{2,3} \otimes {\Tmat}^{\vps}_{4,5} \otimes \dots $.
$\evol$ is hermitian, owing to contraction of a unitary and its conjugate, and is invariant under a shift by
two sites due to ensemble averaging. The matrix $\Tmat^{\vps}_{j,\jp}$ acts only on sites $j$, $\jp$ as
\be   \label{eq:Tdef}  \Tmat^{\vps}_{j,\jp}=\frac{1}{2}(\hat\ident^{\vps}_{j,\jp}  + \hat\swap^{\vps}_{j,\jp})  ,  \ee
where $\hat\swap^{\vps}_{j,\jp} = \frac{1}{2} (\hat\ident^{\vps}_{j,\jp} + \vec{\sigma}^{\vps}_j \cdot \vec{\sigma}^{\vps}_{\jp})$ is the `swap operator'. We note that $\evol$ describes a discrete-time symmetric simple exclusion process (SSEP) for a classical lattice gas \cite{SCHUTZ2001}. Although our original FRUC featured a $U(1)$ symmetry, after Haar averaging and taking $q\to \infty$, $K(t,s)$ exhibits an enlarged $SU(2)$ invariance in the remaining spin-$1/2$ variables; we believe this is specific to the large-$q$ limit. Additionally, as we clarify below, $\evol$ belongs to a family of commuting transfer matrices, unveiling an emergent integrability, and the possibility of computing $K(t, s)$ exactly~\cite{ProsenIntTrot}. 

This model leads to a Thouless time which scales diffusively. To see this, note $\Tmat_{j,j+1} \equiv \hat{\ident}_{j,j+1} - \Ham_{j,j+1}$, where
$\Ham_{j,j+1} = -\frac{1}{4} (\vec\sigma_j \cdot \vec \sigma_{j+1} - \hat{\ident}_{j,j+1})$ describes the spin-$1/2$ Heisenberg ferromagnet. Thus, we can interpret $\Trace [ \evol^t ]$ in \eqref{eq:Ktrace} as a Trotterization of the partition function at inverse temperature $\beta = t$, i.e. $\Trace_s \, [\evol^t] \simeq \Trace_s \, [e^{-t \HXXX}]$, with $\HXXX = \sum_j \Ham_{j,j+1}$. Hence, the behavior of $K(t,s)$ at late times reflects the low-temperature properties of the Heisenberg ferromagnet, $\HXXX$, which has  $(L+1)$-fold degenerate ground states with vanishing energy. Each $S^z$ sector has a unique ground state,
$\ket{S} \equiv (\hat{S}^{-})^{N_{\downarrow}} \ket{\uparrow\ldots\uparrow}$ with $\hat{S}^{\pm} \equiv \sum_{j} \hat \sigma^{\pm}_j$
and $N_{\downarrow} = L/2 -S = L \left( 1/2 - s \right)$. Low-lying excitations above each $\ket{S}$ are \textit{magnons}, i.e. plane-wave superpositions of spin flips 
\begin{equation}
 \ket{S, k} =  \frac{1}{\sqrt{L}}\sum\limits_{j=1}^L e^{\imath j k} \hat{\sigma}^{-}_j \ket{S+1} \;, \quad k = \frac{2\pi p}{L} ~,~~
\end{equation}
characterized by a quadratic dispersion relation $\disp(k) \propto k^2$ at small $k$.
Expanding in $t \gg L^2$, one expects only the lowest energy magnon contributes and 
\be \label{Klarget} 
 \lim_{q \to \infty} \, K(t, s) \underset{~~t \, \gg L^2}{=} \abs{t} \, \left(1 + e^{-\frac{4 \pi^2 t}{L^2}} +\dots ~\right) \, . ~~ \ee
This suggests diffusive scaling of the Thouless time, $\tth \propto L^2$; a similar correspondence with $\HXXX$ was established in \cite{ShenkerRMT} for random unitary circuits with a conserved density, lending support for the generality of this result. However, to investigate the regime $1 \ll t \ll L^2$, we must consider states with extensive numbers of magnons, and many-body effects.

\noindent{\textit{Scaling form.}} We define the function
\be \label{largeLK} \phi(t, s) = -\lim\limits_{L \to \infty}  L^{-1} \, \ln\left[ K \left(t, s \right)/|t|\right] ~, ~~ \ee
which can be computed exactly for any integer $t$, either by solving an infinite set of coupled integral equations---i.e. the Thermodynamic Bethe Ansatz (TBA)~\cite{TakahashiBook,yang1969thermodynamics}---or, perhaps more efficiently, via the `quantum transfer matrix method'~\cite{suzuki2003quantum, klumper_2004}, which requires the solution of an algebraic equation in $2\abs{t}$ variables~\cite{Note2}. While the latter is better suited to calculating $K(t)$ at a \emph{particular} time $t$, 
the former affords analytic insight into behavior at large times. Since the limit $L \to \infty$ implies $\theis \to \infty$, we expand \eqref{largeLK} about large $t$ using TBA,
\be \label{largetphi} \phi(t, s) = - \frac{C}{\sqrt{t}} + \frac{1}{2 (2 s + 1) t} +\dots ~, ~~ \ee
where the constant $C = \zeta(3/2)/\sqrt{4 \pi}$ [$\zeta(z)$ is the Riemann Zeta function] and $q \to \infty$ is taken implicitly. Ignoring the Trotterized structure of $\evol$ and taking $\evol \sim e^{- H_{\rm XXX}}$ relates
\eqref{largetphi} to the low-temperature expansion of the specific heat close to the ferromagnetic ground state of $\HXXX$~\cite{schlottmann1985critical, takahashi1986quantum}.

The form of \eqref{largetphi}  implies diffusive scaling even for $t \ll D L^2$.  From the behavior $\phi(t, s) \sim (D t)^{-1/2}$ it is apparent that the value of $D$ is independent of $s$, a consequence of the emergent $SU(2)$ symmetry at $q \to \infty$. However, the scaling limit relevant for $K(t)$ in the regime $1\ll t \lesssim \tth$ is \emph{distinct} from that recovered from TBA \eqref{largetphi}: the former requires $t, L \to \infty$ with $x\equiv t/L^2$ \emph{fixed}, while the latter requires the thermodynamic limit $L \to \infty$ at fixed $t$. Nevertheless, these results suggest a scaling form
\be  \lim_{t,L \to \infty} \ln \left[K(t, s)/t \right] = \kappa(x, s) \label{eq:Kscalingform} ~.~ \ee
Despite the inherent integrability, exact calculation of $\kappa(x,s)$ is a challenging task. Nevertheless, its asymptotic behavior can be read off from \eqref{Klarget} and \eqref{largetphi}: 
for early times ($x \ll 1$) one inserts \eqref{largeLK} into \eqref{largetphi}; 
for late times ($x \gg 1$) one expands
the log of the right side of \eqref{Klarget}. Thus
\be \label{asykappa}
\kappa(x, s) \underset{x \ll 1}{\sim} C x^{-1/2}\quad {\rm and} \quad \kappa(x, s) \underset{x \gg 1}{\sim} e^{- 4 \pi^2 x}\,.~~\ee
By treating the magnons as non-interacting bosons, using ${\Trace}[ {\evol}^{\, t} \,] \approx \Trace [ e^{- t H}]$, we 
recover~\cite{Note2}
\be \label{fullrangekappa} \kappa \left( x, s \right) = - \sum\limits_{n \neq 0} \ln \left[ 1 - e^{-x D \left( 2 \pi n \right)^2} \right] ~, ~~\ee
which precisely agrees with \eqref{asykappa} if one uses the diffusion constant $D = 1$ associated to the true dispersion \eqref{eq:Mdisp} at small $k$. Although these predictions are obtained for $q \to \infty$, we expect their qualitative features to be valid for generic chaotic many-body systems with conserved charges.
\noindent{\textit{Numerical simulation.}} We now turn to numerical simulation to test \eqref{eq:Kscalingform} in chaotic quantum systems at finite $q$. At $q = 1$, the FRUC considered above exhibits a numerically small diffusion constant that makes it difficult to avoid finite-size effects at the accessible values of $L$. Instead, we use a model adapted from \cite{RUCconTibor}, defined by
\begin{equation}
\Floq = e^{- i t_4 \hat{H}_4} e^{- i t_3 \hat{H}_3} e^{- i t_2 \hat{H}_2} e^{- i t_1 \hat{H}_1} \label{eq:EDFloquet} , 
\end{equation}
where 
\begin{align} 
\hat{H}_1 &=  \sum_j \left(  J^{1}_z \, \hat{\sigma}_j^z \hat{\sigma}_{j+1}^z  + h^{1}_j \, \hat{\sigma}_j^z \right) \notag \\
\hat{H}_3 &=  \sum_j \left(  J^2_z \, \hat{\sigma}_j^z \hat{\sigma}_{j+2}^z  +  h^{2}_j \, \hat{\sigma}_j^z  \right)  \notag \\
\hat{H}_2 &= \hat{H}_4 = J^{\vps}_{xy} \sum_j \left( \hat{\sigma}_j^x \hat{\sigma}_{j+1}^x + \hat{\sigma}_j^y \hat{\sigma}_{j+1}^y \right) \label{eq:NumHs},
\end{align}
with periodic boundary conditions. We take $J^{1}_z = \left( \sqrt{3} + 5 \right)/6$, $J^{2}_z = \sqrt{5}/2$, and $J^{\vps}_{xy} = \left( 2 \sqrt{3} + 3 \right)/7$, with $h^{1,2}_n$ drawn independently from the uniform distribution $\left[ -1.0, 1.0 \right]$ for ensemble averaging. We choose $t_1 = 0.4$, $t_2 = 0.1$, $t_3 = 0.3$, and $t_4 = 0.2$ to avoid time-reversal symmetry around any instant in the period (we check that nearest-neighbor level statistics are CUE). In contrast to a recent study~\cite{Prosen2019mbl}, we did not investigate $K(t)$ at strong disorder, as our concern is with the behavior of ergodic systems.

\begin{figure}[t!]
	\includegraphics[width = 0.95\columnwidth]{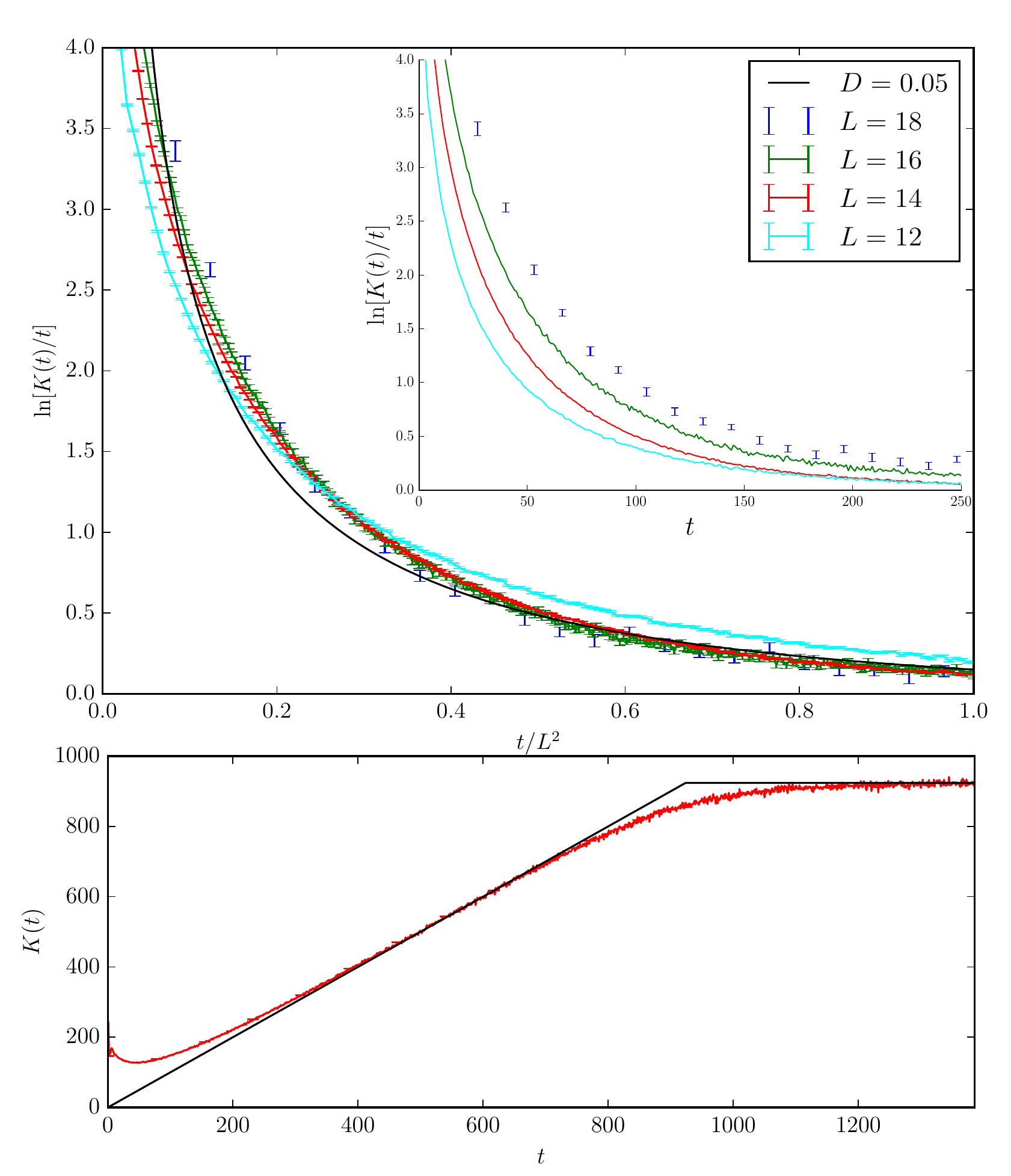} 
	\caption{\label{fig:scaling} Behavior of $K(t)$. Upper figure (main panel): $\ln K \left( t \right)/ t$ vs $t/L^2$. The scaling collapse of data for $L=14$, $16$, and $18$ indicates that the Thouless time $t_{\rm Th} $ is controlled by diffusion; small deviations for $L=12$ are presumably a finite-size effect. The full line is a fit to the scaling function \eqref{fullrangekappa} with $D=0.05$. Inset: same data vs $t$ for comparison.  Lower panel: $K(t)$ for $L=12$;  the small system size narrows the relative extent of the ramp regime $K (t) = t$, highlighting the short-time, pre-RMT behavior. }
\end{figure}

We restrict to half-filling, and measure $K(t)$ for sizes $L= 12,14,16$, each averaged over $\gtrsim 10^4$ disorder configurations \cite{Note2}. Fig.~\ref{fig:scaling} (lower panel)
shows the general behavior of $K(t)$ in a system with a conserved charge, with values much larger than RMT during an initial interval, followed by a linear ramp regime in which $K(t) = \abs{t}$, and finally a plateau for $t \geq \theis$. Fig.~\ref{fig:scaling} (upper panel) shows scaling collapse of $\ln \left[ K \left( t \right) / t \right] $ versus $x = t/L^2$ for different $L$, following \eqref{asykappa}, with a comparison to \eqref{fullrangekappa} with $D$ a fitting parameter, here taken to be $0.05$.
\\
\noindent{\textit{Bethe Ansatz solution.}} We conclude by sketching the analysis of \eqref{eq:Ktrace} using Bethe Ansatz. The computation of $\Trace_s \, [{\evol}^t]$ would be simplified by knowledge of the full eigenspectrum of $\evol$. Following the \emph{coordinate Bethe Ansatz} for $\HXXX$~\cite{gaudin2014bethe, franchini2017introduction}, one seeks multi-magnon eigenfunctions, i.e. plane-waves, along with a scattering matrix describing the exchange of the  excitations' momenta. However, this approach suffers from technical complications due to the circuit construction of $\evol$. A more direct approach is instead based on the equivalent \textit{algebraic Bethe Ansatz} formulation~\cite{korepin1997quantum}. Thus, we introduce the $R$-matrix 
\be \label{Rmatrixdef}  
\Rmat^{\vps}_{a,b}(\lambda) = \frac{\lambda}{\lambda + 2\imath}\hat{\ident}^{\vps}_{ab} + \frac{2\imath}{\lambda + 2\imath}\hat{\swap}^{\vps}_{ab} ~, 
 \ee
which acts on spins-$1/2$ labelled $a$ and $b$. We next introduce the transfer matrix,
which acts jointly on the $L$ physical spins and an auxiliary spin, $a$:
\be \label{TM}
\Transfer (\lambda) \equiv \Rmat^{\vps}_{1,a} (\lambda - \xi_1) \Rmat^{\vps}_{2,a} (\lambda - \xi_{2}) \ldots \Rmat^{\vps}_{ L ,a}(\lambda - \xi_L)\, ,~~~
\ee 
where the \textit{rapidity} $\lambda$ and  \textit{inhomogeneities} $\xi$'s are arbitrary complex numbers. Note that the subscripts in \eqref{TM} label Hilbert spaces, as is customary in the Bethe-Ansatz literature (see also Fig. 4 in \cite{Note2}). Schematically, $\lambda$ parameterizes the quasi-momentum $k(\lambda) = \operatorname{arccot}(\lambda/2) \in (-\pi/2, \pi/2)$ carried by the auxiliary particle while traversing the chain. Alternatively, in the auxiliary space, $\Transfer(\lambda)$ can be written as a $2\times 2$ matrix of operators acting on the \emph{physical} spins,
\be \label{Tmatr}
\Transfer (\lambda) = \left( \begin{array}{ll} \hat{A}(\lambda) & \hat{B}(\lambda) \\ \hat{C}(\lambda) & \hat{D}(\lambda)
\end{array} \right) \;. \ee
This construction is useful because the $R$-matrix in \eqref{Rmatrixdef} satisfies the Yang-Baxter relation, implying a set of algebraic relations between the coefficients in \eqref{Tmatr}, computed at the same inhomogeneities~\cite{
faddeev1996algebraic}; in particular,  setting $\hat{F} \left( \lambda \right) \equiv \hat{A} \left( \lambda \right) + \hat{D} \left( \lambda \right)$, 
one has $[\hat{F} \left( \lambda \right), \hat{F} \left( \lambda^{\prime} \right) ] = 0 \, , ~ \forall ~\lambda, \lambda^{\prime}$. The presence of a one-parameter family of commuting quantities establishes integrability for any choice of $\left\{ \xi_j \right\}$, but only \emph{particular} choices give rise to interesting local models. For instance, the isotropic Heisenberg spin chain is recovered for the homogeneous case, $\xi_j = \imath/2$. However, the brick-wall geometry relevant to $\evol$ is realized via $\xi_j = \imath \, (1 + \left( -1 \right)^j )$, from which it follows~\cite{Note2}~that 
\be \label{eq:MfromF} \evol = \lim\limits_{\delta \to \imath} \hat{F} \left(\imath -\delta \right)^{-1} \hat{F} \left( \imath +\delta \right) \, , \ee
where the limit is needed to account for the non-invertibility of $\Tmat^{\vps}_{j,\jp}$ in \eqref{eq:Tdef}. 
The common eigenstates of the conserved quantities $\hat{F}(\lambda)$ (and thereby $\evol$) can be obtained via algebraic properties, from which one can interpret $\hat B(\lambda)$ as an effective magnon creation operator on the vacuum $\ket{S}$, which decreases $\hat{S}^z$ by one. Thus $\evol$ has eigenstates
\be \label{eq:Meigs} \evol \ket{\lambda_1,\ldots,\lambda_N}_S = e^{- \sum_j \disp(\lambda_j)} \ket{\lambda_1,\ldots,\lambda_N} \, , ~~
\ee 
where $\ket{\lambda_1,\ldots,\lambda_N}_S = \hat{B} (\lambda_1)\ldots \hat{B} (\lambda_N) \ket{S}$. The integer $N$ encodes the magnetization eigenvalue via
$\hat S^z \ket{\lambda_1,\ldots,\lambda_N}_S = (S-N)\ket{\lambda_1,\ldots,\lambda_N}$.  
Due to interactions, the parameters $\lambda_1,\ldots,\lambda_N$ are not free, but satisfy
\be \label{BASEP}
\left(\frac{\lambda_j + 2\imath}{\lambda_j - 2\imath}\right)^{L/2}
= 
\prod_{\substack{\jp = 1 \\ \jp \neq j}}^{N} \left(\frac{\lambda_j - \lambda_{\jp} + 2\imath }{\lambda_j - \lambda_{\jp} - 2\imath} \right) ~, ~~
\ee
the solution of which provides the full spectrum of $\evol$. 

The dispersion relation in \eqref{eq:Meigs} is given by
\be \label{eq:Mdisp} \disp( \lambda ) \equiv - 2 \ln \cos k(\lambda). \ee 
At small $k$, a quadratic dispersion relation is recovered, as is expected since the  
discrepancies between $\evol^t$ and $e^{-t\HXXX}$ become irrelevant at long wavelengths. 
Although (\ref{eq:Meigs},\ref{BASEP}) simplify substantially the evaluation of \eqref{eq:Ktrace}, the main bottleneck remains the exponential growth in $L$ of the Hilbert space dimension. 
However, in the thermodynamic limit $L, S \to \infty$ with $s=S/L $ fixed, the solutions of \eqref{BASEP} acquire a simple structure \cite{Note2}, and the function $\phi(t,s)$ in \eqref{largeLK} can be computed analogously to thermodynamic quantities in integrable spin chains.

\noindent{\textit{Discussion.}} We have presented analytical and numerical evidence showing the significance of the Thouless time $\tth=L^2/D$ for spectral correlations in systems with a conserved charge.  These results are consistent with a scaling form $ K ( t) \sim \abs{t} \exp \left[ \kappa (t/L^2) \right]$. We believe this form to be generic in describing the onset of chaos in quantum systems with a conserved quantity. We also provide a form \eqref{fullrangekappa} of the scaling function $\kappa$ \eqref{eq:Kscalingform} by neglecting the interactions between magnons in our FRUC, which is in good agreement with our numerical simulations. The question of the exactness and universality of \eqref{fullrangekappa} is an interesting topic for future study.
Another interesting perspective for future work is the study of systems supporting non-Abelian symmetries, where the interplay of different conserved quantities can give rise to anomalous transport~\cite{nlfhHuse,PhysRevLett.122.127202,anomalousDN}.  Finally, while this manuscript was in preparation, related numerical results were presented in Ref.~\citenum{Prosen2019mbl} for the scaling of $\tth$ with $L^{-2}$ in a Hamiltonian (rather than Floquet) model.

\noindent{\textit{Acknowledgments.}} 
We thank A.~Nahum, T.~Rakovszky, and B.~Ware for useful discussions. This research was supported in part by the National Science Foundation via Grants DGE-1321846 (Graduate Research Fellowship Program) and DMR-1455366 (AJF), by EPSRC Grant No.
EP/N01930X/1 (JTC) and the European Unions Horizon 2020 research and innovation programme under the Marie Sklodowska-Curie Grant Agreement No.~794750 (A.D.L.).
We also acknowledge the hospitality of the Kavli Institute for Theoretical Physics at the University of California, Santa Barbara (supported by NSF Grant PHY-1748958), and of 
the University of Massachusetts Amherst (AJF), where parts of this work were completed.

\bibliography{RUFC}

\begin{thebibliography}{56}%
\makeatletter
\providecommand \@ifxundefined [1]{%
 \@ifx{#1\undefined}
}%
\providecommand \@ifnum [1]{%
 \ifnum #1\expandafter \@firstoftwo
 \else \expandafter \@secondoftwo
 \fi
}%
\providecommand \@ifx [1]{%
 \ifx #1\expandafter \@firstoftwo
 \else \expandafter \@secondoftwo
 \fi
}%
\providecommand \natexlab [1]{#1}%
\providecommand \enquote  [1]{``#1''}%
\providecommand \bibnamefont  [1]{#1}%
\providecommand \bibfnamefont [1]{#1}%
\providecommand \citenamefont [1]{#1}%
\providecommand \href@noop [0]{\@secondoftwo}%
\providecommand \href [0]{\begingroup \@sanitize@url \@href}%
\providecommand \@href[1]{\@@startlink{#1}\@@href}%
\providecommand \@@href[1]{\endgroup#1\@@endlink}%
\providecommand \@sanitize@url [0]{\catcode `\\12\catcode `\$12\catcode
  `\&12\catcode `\#12\catcode `\^12\catcode `\_12\catcode `\%12\relax}%
\providecommand \@@startlink[1]{}%
\providecommand \@@endlink[0]{}%
\providecommand \url  [0]{\begingroup\@sanitize@url \@url }%
\providecommand \@url [1]{\endgroup\@href {#1}{\urlprefix }}%
\providecommand \urlprefix  [0]{URL }%
\providecommand \Eprint [0]{\href }%
\providecommand \doibase [0]{http://dx.doi.org/}%
\providecommand \selectlanguage [0]{\@gobble}%
\providecommand \bibinfo  [0]{\@secondoftwo}%
\providecommand \bibfield  [0]{\@secondoftwo}%
\providecommand \translation [1]{[#1]}%
\providecommand \BibitemOpen [0]{}%
\providecommand \bibitemStop [0]{}%
\providecommand \bibitemNoStop [0]{.\EOS\space}%
\providecommand \EOS [0]{\spacefactor3000\relax}%
\providecommand \BibitemShut  [1]{\csname bibitem#1\endcsname}%
\let\auto@bib@innerbib\@empty
\bibitem [{\citenamefont {D'Alessio}\ \emph {et~al.}(2016)\citenamefont
  {D'Alessio}, \citenamefont {Kafri}, \citenamefont {Polkovnikov},\ and\
  \citenamefont {Rigol}}]{Rigol}%
  \BibitemOpen
  \bibfield  {author} {\bibinfo {author} {\bibfnamefont {L.}~\bibnamefont
  {D'Alessio}}, \bibinfo {author} {\bibfnamefont {Y.}~\bibnamefont {Kafri}},
  \bibinfo {author} {\bibfnamefont {A.}~\bibnamefont {Polkovnikov}}, \ and\
  \bibinfo {author} {\bibfnamefont {M.}~\bibnamefont {Rigol}},\ }\href
  {\doibase 10.1080/00018732.2016.1198134} {\bibfield  {journal} {\bibinfo
  {journal} {Advances in Physics}\ }\textbf {\bibinfo {volume} {65}},\ \bibinfo
  {pages} {239} (\bibinfo {year} {2016})}\BibitemShut {NoStop}%
\bibitem [{\citenamefont {Bloch}\ \emph {et~al.}(2008)\citenamefont {Bloch},
  \citenamefont {Dalibard},\ and\ \citenamefont {Zwerger}}]{RevModPhys.80.885}%
  \BibitemOpen
  \bibfield  {author} {\bibinfo {author} {\bibfnamefont {I.}~\bibnamefont
  {Bloch}}, \bibinfo {author} {\bibfnamefont {J.}~\bibnamefont {Dalibard}}, \
  and\ \bibinfo {author} {\bibfnamefont {W.}~\bibnamefont {Zwerger}},\ }\href
  {\doibase 10.1103/RevModPhys.80.885} {\bibfield  {journal} {\bibinfo
  {journal} {Rev. Mod. Phys.}\ }\textbf {\bibinfo {volume} {80}},\ \bibinfo
  {pages} {885} (\bibinfo {year} {2008})}\BibitemShut {NoStop}%
\bibitem [{\citenamefont {Bloch}\ \emph {et~al.}(2012)\citenamefont {Bloch},
  \citenamefont {Dalibard},\ and\ \citenamefont
  {Nascimbene}}]{bloch2012quantum}%
  \BibitemOpen
  \bibfield  {author} {\bibinfo {author} {\bibfnamefont {I.}~\bibnamefont
  {Bloch}}, \bibinfo {author} {\bibfnamefont {J.}~\bibnamefont {Dalibard}}, \
  and\ \bibinfo {author} {\bibfnamefont {S.}~\bibnamefont {Nascimbene}},\
  }\href@noop {} {\bibfield  {journal} {\bibinfo  {journal} {Nature Physics}\
  }\textbf {\bibinfo {volume} {8}},\ \bibinfo {pages} {267} (\bibinfo {year}
  {2012})}\BibitemShut {NoStop}%
\bibitem [{\citenamefont {Deutsch}(1991)}]{ETH1}%
  \BibitemOpen
  \bibfield  {author} {\bibinfo {author} {\bibfnamefont {J.~M.}\ \bibnamefont
  {Deutsch}},\ }\href {\doibase 10.1103/PhysRevA.43.2046} {\bibfield  {journal}
  {\bibinfo  {journal} {Phys. Rev. A}\ }\textbf {\bibinfo {volume} {43}},\
  \bibinfo {pages} {2046} (\bibinfo {year} {1991})}\BibitemShut {NoStop}%
\bibitem [{\citenamefont {Srednicki}(1994)}]{ETH2}%
  \BibitemOpen
  \bibfield  {author} {\bibinfo {author} {\bibfnamefont {M.}~\bibnamefont
  {Srednicki}},\ }\href {\doibase 10.1103/PhysRevE.50.888} {\bibfield
  {journal} {\bibinfo  {journal} {Phys. Rev. E}\ }\textbf {\bibinfo {volume}
  {50}},\ \bibinfo {pages} {888} (\bibinfo {year} {1994})}\BibitemShut
  {NoStop}%
\bibitem [{\citenamefont {Rigol}\ \emph {et~al.}(2008)\citenamefont {Rigol},
  \citenamefont {Dunjko},\ and\ \citenamefont {Olshanii}}]{Rigol2008kq}%
  \BibitemOpen
  \bibfield  {author} {\bibinfo {author} {\bibfnamefont {M.}~\bibnamefont
  {Rigol}}, \bibinfo {author} {\bibfnamefont {V.}~\bibnamefont {Dunjko}}, \
  and\ \bibinfo {author} {\bibfnamefont {M.}~\bibnamefont {Olshanii}},\ }\href
  {http://dx.doi.org/10.1038/nature06838} {\bibfield  {journal} {\bibinfo
  {journal} {Nature}\ }\textbf {\bibinfo {volume} {452}},\ \bibinfo {pages}
  {854} (\bibinfo {year} {2008})}\BibitemShut {NoStop}%
\bibitem [{\citenamefont {Bohigas}\ \emph {et~al.}(1984)\citenamefont
  {Bohigas}, \citenamefont {Giannoni},\ and\ \citenamefont
  {Schmit}}]{bohigas1984characterization}%
  \BibitemOpen
  \bibfield  {author} {\bibinfo {author} {\bibfnamefont {O.}~\bibnamefont
  {Bohigas}}, \bibinfo {author} {\bibfnamefont {M.-J.}\ \bibnamefont
  {Giannoni}}, \ and\ \bibinfo {author} {\bibfnamefont {C.}~\bibnamefont
  {Schmit}},\ }\href@noop {} {\bibfield  {journal} {\bibinfo  {journal} {Phys.
  Rev. Lett.}\ }\textbf {\bibinfo {volume} {52}},\ \bibinfo {pages} {1}
  (\bibinfo {year} {1984})}\BibitemShut {NoStop}%
\bibitem [{\citenamefont {Guhr}\ \emph {et~al.}(1998)\citenamefont {Guhr},
  \citenamefont {M{\"u}ller-Groeling},\ and\ \citenamefont
  {Weidenm{\"u}ller}}]{guhr1998random}%
  \BibitemOpen
  \bibfield  {author} {\bibinfo {author} {\bibfnamefont {T.}~\bibnamefont
  {Guhr}}, \bibinfo {author} {\bibfnamefont {A.}~\bibnamefont
  {M{\"u}ller-Groeling}}, \ and\ \bibinfo {author} {\bibfnamefont {H.~A.}\
  \bibnamefont {Weidenm{\"u}ller}},\ }\href@noop {} {\bibfield  {journal}
  {\bibinfo  {journal} {Physics Reports}\ }\textbf {\bibinfo {volume} {299}},\
  \bibinfo {pages} {189} (\bibinfo {year} {1998})}\BibitemShut {NoStop}%
\bibitem [{\citenamefont {Berry}(1977)}]{berry1977mv}%
  \BibitemOpen
  \bibfield  {author} {\bibinfo {author} {\bibfnamefont {M.}~\bibnamefont
  {Berry}},\ }\href@noop {} {\bibfield  {journal} {\bibinfo  {journal} {J.
  Phys. A}\ }\textbf {\bibinfo {volume} {10}},\ \bibinfo {pages} {2083}
  (\bibinfo {year} {1977})}\BibitemShut {NoStop}%
\bibitem [{\citenamefont {Borgonovi}\ \emph {et~al.}(2016)\citenamefont
  {Borgonovi}, \citenamefont {Izrailev}, \citenamefont {Santos},\ and\
  \citenamefont {Zelevinsky}}]{borgonovi2016quantum}%
  \BibitemOpen
  \bibfield  {author} {\bibinfo {author} {\bibfnamefont {F.}~\bibnamefont
  {Borgonovi}}, \bibinfo {author} {\bibfnamefont {F.~M.}\ \bibnamefont
  {Izrailev}}, \bibinfo {author} {\bibfnamefont {L.~F.}\ \bibnamefont
  {Santos}}, \ and\ \bibinfo {author} {\bibfnamefont {V.~G.}\ \bibnamefont
  {Zelevinsky}},\ }\href@noop {} {\bibfield  {journal} {\bibinfo  {journal}
  {Physics Reports}\ }\textbf {\bibinfo {volume} {626}},\ \bibinfo {pages} {1}
  (\bibinfo {year} {2016})}\BibitemShut {NoStop}%
\bibitem [{\citenamefont {Rigol}(2009)}]{rigol2009breakdown}%
  \BibitemOpen
  \bibfield  {author} {\bibinfo {author} {\bibfnamefont {M.}~\bibnamefont
  {Rigol}},\ }\href@noop {} {\bibfield  {journal} {\bibinfo  {journal}
  {Physical Review Letters}\ }\textbf {\bibinfo {volume} {103}},\ \bibinfo
  {pages} {100403} (\bibinfo {year} {2009})}\BibitemShut {NoStop}%
\bibitem [{\citenamefont {Santos}\ and\ \citenamefont
  {Rigol}(2010)}]{santos2010onset}%
  \BibitemOpen
  \bibfield  {author} {\bibinfo {author} {\bibfnamefont {L.~F.}\ \bibnamefont
  {Santos}}\ and\ \bibinfo {author} {\bibfnamefont {M.}~\bibnamefont {Rigol}},\
  }\href@noop {} {\bibfield  {journal} {\bibinfo  {journal} {Physical Review
  E}\ }\textbf {\bibinfo {volume} {81}},\ \bibinfo {pages} {036206} (\bibinfo
  {year} {2010})}\BibitemShut {NoStop}%
\bibitem [{\citenamefont {Biroli}\ \emph {et~al.}(2010)\citenamefont {Biroli},
  \citenamefont {Kollath},\ and\ \citenamefont
  {L{\"a}uchli}}]{biroli2010effect}%
  \BibitemOpen
  \bibfield  {author} {\bibinfo {author} {\bibfnamefont {G.}~\bibnamefont
  {Biroli}}, \bibinfo {author} {\bibfnamefont {C.}~\bibnamefont {Kollath}}, \
  and\ \bibinfo {author} {\bibfnamefont {A.~M.}\ \bibnamefont {L{\"a}uchli}},\
  }\href@noop {} {\bibfield  {journal} {\bibinfo  {journal} {Physical Review
  Letters}\ }\textbf {\bibinfo {volume} {105}},\ \bibinfo {pages} {250401}
  (\bibinfo {year} {2010})}\BibitemShut {NoStop}%
\bibitem [{\citenamefont {Pal}\ and\ \citenamefont {Huse}(2010)}]{pal2010many}%
  \BibitemOpen
  \bibfield  {author} {\bibinfo {author} {\bibfnamefont {A.}~\bibnamefont
  {Pal}}\ and\ \bibinfo {author} {\bibfnamefont {D.~A.}\ \bibnamefont {Huse}},\
  }\href@noop {} {\bibfield  {journal} {\bibinfo  {journal} {Phys. Rev. B}\
  }\textbf {\bibinfo {volume} {82}},\ \bibinfo {pages} {174411} (\bibinfo
  {year} {2010})}\BibitemShut {NoStop}%
\bibitem [{\citenamefont {Alba}(2015)}]{alba2015eigenstate}%
  \BibitemOpen
  \bibfield  {author} {\bibinfo {author} {\bibfnamefont {V.}~\bibnamefont
  {Alba}},\ }\href@noop {} {\bibfield  {journal} {\bibinfo  {journal} {Phys.
  Rev. B}\ }\textbf {\bibinfo {volume} {91}},\ \bibinfo {pages} {155123}
  (\bibinfo {year} {2015})}\BibitemShut {NoStop}%
\bibitem [{\citenamefont {Rigol}\ and\ \citenamefont
  {Srednicki}(2012)}]{rigol2012alternatives}%
  \BibitemOpen
  \bibfield  {author} {\bibinfo {author} {\bibfnamefont {M.}~\bibnamefont
  {Rigol}}\ and\ \bibinfo {author} {\bibfnamefont {M.}~\bibnamefont
  {Srednicki}},\ }\href@noop {} {\bibfield  {journal} {\bibinfo  {journal}
  {Phys. Rev. Lett.}\ }\textbf {\bibinfo {volume} {108}},\ \bibinfo {pages}
  {110601} (\bibinfo {year} {2012})}\BibitemShut {NoStop}%
\bibitem [{\citenamefont {Luitz}\ and\ \citenamefont
  {Lev}(2016)}]{luitz2016anomalous}%
  \BibitemOpen
  \bibfield  {author} {\bibinfo {author} {\bibfnamefont {D.~J.}\ \bibnamefont
  {Luitz}}\ and\ \bibinfo {author} {\bibfnamefont {Y.~B.}\ \bibnamefont
  {Lev}},\ }\href@noop {} {\bibfield  {journal} {\bibinfo  {journal} {Physical
  review letters}\ }\textbf {\bibinfo {volume} {117}},\ \bibinfo {pages}
  {170404} (\bibinfo {year} {2016})}\BibitemShut {NoStop}%
\bibitem [{\citenamefont {{Dymarsky}}(2018)}]{dymarsky}%
  \BibitemOpen
  \bibfield  {author} {\bibinfo {author} {\bibfnamefont {A.}~\bibnamefont
  {{Dymarsky}}},\ }\href@noop {} {\bibfield  {journal} {\bibinfo  {journal}
  {arXiv preprint arXiv:1804.08626}\ } (\bibinfo {year} {2018})}\BibitemShut
  {NoStop}%
\bibitem [{\citenamefont {Foini}\ and\ \citenamefont
  {Kurchan}(2019{\natexlab{a}})}]{foini2019eigenstate}%
  \BibitemOpen
  \bibfield  {author} {\bibinfo {author} {\bibfnamefont {L.}~\bibnamefont
  {Foini}}\ and\ \bibinfo {author} {\bibfnamefont {J.}~\bibnamefont
  {Kurchan}},\ }\href@noop {} {\bibfield  {journal} {\bibinfo  {journal}
  {Physical Review E}\ }\textbf {\bibinfo {volume} {99}},\ \bibinfo {pages}
  {042139} (\bibinfo {year} {2019}{\natexlab{a}})}\BibitemShut {NoStop}%
\bibitem [{\citenamefont {Foini}\ and\ \citenamefont
  {Kurchan}(2019{\natexlab{b}})}]{foini2019rotational}%
  \BibitemOpen
  \bibfield  {author} {\bibinfo {author} {\bibfnamefont {L.}~\bibnamefont
  {Foini}}\ and\ \bibinfo {author} {\bibfnamefont {J.}~\bibnamefont
  {Kurchan}},\ }\href@noop {} {\bibfield  {journal} {\bibinfo  {journal} {arXiv
  preprint arXiv:1906.01522}\ } (\bibinfo {year}
  {2019}{\natexlab{b}})}\BibitemShut {NoStop}%
\bibitem [{\citenamefont {{Chan}}\ \emph {et~al.}(2018)\citenamefont {{Chan}},
  \citenamefont {{De Luca}},\ and\ \citenamefont {{Chalker}}}]{CDLC3}%
  \BibitemOpen
  \bibfield  {author} {\bibinfo {author} {\bibfnamefont {A.}~\bibnamefont
  {{Chan}}}, \bibinfo {author} {\bibfnamefont {A.}~\bibnamefont {{De Luca}}}, \
  and\ \bibinfo {author} {\bibfnamefont {J.~T.}\ \bibnamefont {{Chalker}}},\
  }\href@noop {} {\bibfield  {journal} {\bibinfo  {journal} {arXiv e-prints}\
  ,\ \bibinfo {eid} {arXiv:1810.11014}} (\bibinfo {year} {2018})},\ \Eprint
  {http://arxiv.org/abs/1810.11014} {arXiv:1810.11014 [cond-mat.stat-mech]}
  \BibitemShut {NoStop}%
\bibitem [{\citenamefont {Brandino}\ \emph {et~al.}(2011)\citenamefont
  {Brandino}, \citenamefont {De~Luca}, \citenamefont {Konik},\ and\
  \citenamefont {Mussardo}}]{brandino2011quench}%
  \BibitemOpen
  \bibfield  {author} {\bibinfo {author} {\bibfnamefont {G.}~\bibnamefont
  {Brandino}}, \bibinfo {author} {\bibfnamefont {A.}~\bibnamefont {De~Luca}},
  \bibinfo {author} {\bibfnamefont {R.}~\bibnamefont {Konik}}, \ and\ \bibinfo
  {author} {\bibfnamefont {G.}~\bibnamefont {Mussardo}},\ }\href@noop {}
  {\bibfield  {journal} {\bibinfo  {journal} {Physical Review B, 2012}\
  }\textbf {\bibinfo {volume} {85}},\ \bibinfo {pages} {214435} (\bibinfo
  {year} {2011})}\BibitemShut {NoStop}%
\bibitem [{\citenamefont {Beugeling}\ \emph {et~al.}(2015)\citenamefont
  {Beugeling}, \citenamefont {Moessner},\ and\ \citenamefont
  {Haque}}]{MoessnerHaque}%
  \BibitemOpen
  \bibfield  {author} {\bibinfo {author} {\bibfnamefont {W.}~\bibnamefont
  {Beugeling}}, \bibinfo {author} {\bibfnamefont {R.}~\bibnamefont {Moessner}},
  \ and\ \bibinfo {author} {\bibfnamefont {M.}~\bibnamefont {Haque}},\ }\href
  {\doibase 10.1103/PhysRevE.91.012144} {\bibfield  {journal} {\bibinfo
  {journal} {Phys. Rev. E}\ }\textbf {\bibinfo {volume} {91}},\ \bibinfo
  {pages} {012144} (\bibinfo {year} {2015})}\BibitemShut {NoStop}%
\bibitem [{\citenamefont {Kim}\ \emph {et~al.}(2014)\citenamefont {Kim},
  \citenamefont {Ikeda},\ and\ \citenamefont {Huse}}]{kim2014testing}%
  \BibitemOpen
  \bibfield  {author} {\bibinfo {author} {\bibfnamefont {H.}~\bibnamefont
  {Kim}}, \bibinfo {author} {\bibfnamefont {T.~N.}\ \bibnamefont {Ikeda}}, \
  and\ \bibinfo {author} {\bibfnamefont {D.~A.}\ \bibnamefont {Huse}},\
  }\href@noop {} {\bibfield  {journal} {\bibinfo  {journal} {Physical Review
  E}\ }\textbf {\bibinfo {volume} {90}},\ \bibinfo {pages} {052105} (\bibinfo
  {year} {2014})}\BibitemShut {NoStop}%
\bibitem [{\citenamefont {Nahum}\ \emph {et~al.}(2017)\citenamefont {Nahum},
  \citenamefont {Ruhman}, \citenamefont {Vijay},\ and\ \citenamefont
  {Haah}}]{NahumRUC1}%
  \BibitemOpen
  \bibfield  {author} {\bibinfo {author} {\bibfnamefont {A.}~\bibnamefont
  {Nahum}}, \bibinfo {author} {\bibfnamefont {J.}~\bibnamefont {Ruhman}},
  \bibinfo {author} {\bibfnamefont {S.}~\bibnamefont {Vijay}}, \ and\ \bibinfo
  {author} {\bibfnamefont {J.}~\bibnamefont {Haah}},\ }\href {\doibase
  10.1103/PhysRevX.7.031016} {\bibfield  {journal} {\bibinfo  {journal} {Phys.
  Rev. X}\ }\textbf {\bibinfo {volume} {7}},\ \bibinfo {pages} {031016}
  (\bibinfo {year} {2017})}\BibitemShut {NoStop}%
\bibitem [{\citenamefont {von Keyserlingk}\ \emph {et~al.}(2018)\citenamefont
  {von Keyserlingk}, \citenamefont {Rakovszky}, \citenamefont {Pollmann},\ and\
  \citenamefont {Sondhi}}]{RUCNCTibor}%
  \BibitemOpen
  \bibfield  {author} {\bibinfo {author} {\bibfnamefont {C.~W.}\ \bibnamefont
  {von Keyserlingk}}, \bibinfo {author} {\bibfnamefont {T.}~\bibnamefont
  {Rakovszky}}, \bibinfo {author} {\bibfnamefont {F.}~\bibnamefont {Pollmann}},
  \ and\ \bibinfo {author} {\bibfnamefont {S.~L.}\ \bibnamefont {Sondhi}},\
  }\href {\doibase 10.1103/PhysRevX.8.021013} {\bibfield  {journal} {\bibinfo
  {journal} {Phys. Rev. X}\ }\textbf {\bibinfo {volume} {8}},\ \bibinfo {pages}
  {021013} (\bibinfo {year} {2018})}\BibitemShut {NoStop}%
\bibitem [{\citenamefont {Chan}\ \emph
  {et~al.}(2018{\natexlab{a}})\citenamefont {Chan}, \citenamefont {De~Luca},\
  and\ \citenamefont {Chalker}}]{CDLC1}%
  \BibitemOpen
  \bibfield  {author} {\bibinfo {author} {\bibfnamefont {A.}~\bibnamefont
  {Chan}}, \bibinfo {author} {\bibfnamefont {A.}~\bibnamefont {De~Luca}}, \
  and\ \bibinfo {author} {\bibfnamefont {J.~T.}\ \bibnamefont {Chalker}},\
  }\href {\doibase 10.1103/PhysRevX.8.041019} {\bibfield  {journal} {\bibinfo
  {journal} {Phys. Rev. X}\ }\textbf {\bibinfo {volume} {8}},\ \bibinfo {pages}
  {041019} (\bibinfo {year} {2018}{\natexlab{a}})}\BibitemShut {NoStop}%
\bibitem [{\citenamefont {Chan}\ \emph
  {et~al.}(2018{\natexlab{b}})\citenamefont {Chan}, \citenamefont {De~Luca},\
  and\ \citenamefont {Chalker}}]{CDLC2}%
  \BibitemOpen
  \bibfield  {author} {\bibinfo {author} {\bibfnamefont {A.}~\bibnamefont
  {Chan}}, \bibinfo {author} {\bibfnamefont {A.}~\bibnamefont {De~Luca}}, \
  and\ \bibinfo {author} {\bibfnamefont {J.~T.}\ \bibnamefont {Chalker}},\
  }\href {\doibase 10.1103/PhysRevLett.121.060601} {\bibfield  {journal}
  {\bibinfo  {journal} {Phys. Rev. Lett.}\ }\textbf {\bibinfo {volume} {121}},\
  \bibinfo {pages} {060601} (\bibinfo {year} {2018}{\natexlab{b}})}\BibitemShut
  {NoStop}%
\bibitem [{\citenamefont {Rakovszky}\ \emph {et~al.}(2018)\citenamefont
  {Rakovszky}, \citenamefont {Pollmann},\ and\ \citenamefont {von
  Keyserlingk}}]{RUCconTibor}%
  \BibitemOpen
  \bibfield  {author} {\bibinfo {author} {\bibfnamefont {T.}~\bibnamefont
  {Rakovszky}}, \bibinfo {author} {\bibfnamefont {F.}~\bibnamefont {Pollmann}},
  \ and\ \bibinfo {author} {\bibfnamefont {C.~W.}\ \bibnamefont {von
  Keyserlingk}},\ }\href {\doibase 10.1103/PhysRevX.8.031058} {\bibfield
  {journal} {\bibinfo  {journal} {Phys. Rev. X}\ }\textbf {\bibinfo {volume}
  {8}},\ \bibinfo {pages} {031058} (\bibinfo {year} {2018})}\BibitemShut
  {NoStop}%
\bibitem [{\citenamefont {Khemani}\ \emph {et~al.}(2018)\citenamefont
  {Khemani}, \citenamefont {Vishwanath},\ and\ \citenamefont
  {Huse}}]{RUCconVedika}%
  \BibitemOpen
  \bibfield  {author} {\bibinfo {author} {\bibfnamefont {V.}~\bibnamefont
  {Khemani}}, \bibinfo {author} {\bibfnamefont {A.}~\bibnamefont {Vishwanath}},
  \ and\ \bibinfo {author} {\bibfnamefont {D.~A.}\ \bibnamefont {Huse}},\
  }\href {\doibase 10.1103/PhysRevX.8.031057} {\bibfield  {journal} {\bibinfo
  {journal} {Phys. Rev. X}\ }\textbf {\bibinfo {volume} {8}},\ \bibinfo {pages}
  {031057} (\bibinfo {year} {2018})}\BibitemShut {NoStop}%
\bibitem [{\citenamefont {S\"underhauf}\ \emph {et~al.}(2018)\citenamefont
  {S\"underhauf}, \citenamefont {P\'erez-Garc\'{\i}a}, \citenamefont {Huse},
  \citenamefont {Schuch},\ and\ \citenamefont {Cirac}}]{CiracRUFC}%
  \BibitemOpen
  \bibfield  {author} {\bibinfo {author} {\bibfnamefont {C.}~\bibnamefont
  {S\"underhauf}}, \bibinfo {author} {\bibfnamefont {D.}~\bibnamefont
  {P\'erez-Garc\'{\i}a}}, \bibinfo {author} {\bibfnamefont {D.~A.}\
  \bibnamefont {Huse}}, \bibinfo {author} {\bibfnamefont {N.}~\bibnamefont
  {Schuch}}, \ and\ \bibinfo {author} {\bibfnamefont {J.~I.}\ \bibnamefont
  {Cirac}},\ }\href {\doibase 10.1103/PhysRevB.98.134204} {\bibfield  {journal}
  {\bibinfo  {journal} {Phys. Rev. B}\ }\textbf {\bibinfo {volume} {98}},\
  \bibinfo {pages} {134204} (\bibinfo {year} {2018})}\BibitemShut {NoStop}%
\bibitem [{\citenamefont {Kos}\ \emph {et~al.}(2018)\citenamefont {Kos},
  \citenamefont {Ljubotina},\ and\ \citenamefont {Prosen}}]{ProsenRMTChaosPRX}%
  \BibitemOpen
  \bibfield  {author} {\bibinfo {author} {\bibfnamefont {P.}~\bibnamefont
  {Kos}}, \bibinfo {author} {\bibfnamefont {M.}~\bibnamefont {Ljubotina}}, \
  and\ \bibinfo {author} {\bibfnamefont {T.~c.~v.}\ \bibnamefont {Prosen}},\
  }\href {\doibase 10.1103/PhysRevX.8.021062} {\bibfield  {journal} {\bibinfo
  {journal} {Phys. Rev. X}\ }\textbf {\bibinfo {volume} {8}},\ \bibinfo {pages}
  {021062} (\bibinfo {year} {2018})}\BibitemShut {NoStop}%
\bibitem [{\citenamefont {Gharibyan}\ \emph {et~al.}(2018)\citenamefont
  {Gharibyan}, \citenamefont {Hanada}, \citenamefont {Shenker},\ and\
  \citenamefont {Tezuka}}]{ShenkerRMT}%
  \BibitemOpen
  \bibfield  {author} {\bibinfo {author} {\bibfnamefont {H.}~\bibnamefont
  {Gharibyan}}, \bibinfo {author} {\bibfnamefont {M.}~\bibnamefont {Hanada}},
  \bibinfo {author} {\bibfnamefont {S.~H.}\ \bibnamefont {Shenker}}, \ and\
  \bibinfo {author} {\bibfnamefont {M.}~\bibnamefont {Tezuka}},\ }\href
  {\doibase 10.1007/JHEP07(2018)124} {\bibfield  {journal} {\bibinfo  {journal}
  {Journal of High Energy Physics}\ }\textbf {\bibinfo {volume} {2018}},\
  \bibinfo {pages} {124} (\bibinfo {year} {2018})}\BibitemShut {NoStop}%
\bibitem [{\citenamefont {Bertini}\ \emph {et~al.}(2018)\citenamefont
  {Bertini}, \citenamefont {Kos},\ and\ \citenamefont
  {Prosen}}]{bertini2018exact}%
  \BibitemOpen
  \bibfield  {author} {\bibinfo {author} {\bibfnamefont {B.}~\bibnamefont
  {Bertini}}, \bibinfo {author} {\bibfnamefont {P.}~\bibnamefont {Kos}}, \ and\
  \bibinfo {author} {\bibfnamefont {T.}~\bibnamefont {Prosen}},\ }\href@noop {}
  {\bibfield  {journal} {\bibinfo  {journal} {Physical review letters}\
  }\textbf {\bibinfo {volume} {121}},\ \bibinfo {pages} {264101} (\bibinfo
  {year} {2018})}\BibitemShut {NoStop}%
\bibitem [{\citenamefont {Thouless}(1977)}]{ThoulessPRL}%
  \BibitemOpen
  \bibfield  {author} {\bibinfo {author} {\bibfnamefont {D.~J.}\ \bibnamefont
  {Thouless}},\ }\href {\doibase 10.1103/PhysRevLett.39.1167} {\bibfield
  {journal} {\bibinfo  {journal} {Phys. Rev. Lett.}\ }\textbf {\bibinfo
  {volume} {39}},\ \bibinfo {pages} {1167} (\bibinfo {year}
  {1977})}\BibitemShut {NoStop}%
\bibitem [{\citenamefont {Altshuler}\ and\ \citenamefont
  {Shklovskii}(1986)}]{AltshulerShklovskii}%
  \BibitemOpen
  \bibfield  {author} {\bibinfo {author} {\bibfnamefont {B.~L.}\ \bibnamefont
  {Altshuler}}\ and\ \bibinfo {author} {\bibfnamefont {B.~I.}\ \bibnamefont
  {Shklovskii}},\ }\href {http://www.jetp.ac.ru/cgi-bin/dn/e_064_01_0127.pdf}
  {\bibfield  {journal} {\bibinfo  {journal} {JETP}\ }\textbf {\bibinfo
  {volume} {64}},\ \bibinfo {pages} {127} (\bibinfo {year} {1986})}\BibitemShut
  {NoStop}%
\bibitem [{\citenamefont {Braun}\ \emph {et~al.}(2019)\citenamefont {Braun},
  \citenamefont {Waltner}, \citenamefont {Akila}, \citenamefont {Gutkin},\ and\
  \citenamefont {Guhr}}]{GuhrArxiv}%
  \BibitemOpen
  \bibfield  {author} {\bibinfo {author} {\bibfnamefont {P.}~\bibnamefont
  {Braun}}, \bibinfo {author} {\bibfnamefont {D.}~\bibnamefont {Waltner}},
  \bibinfo {author} {\bibfnamefont {M.}~\bibnamefont {Akila}}, \bibinfo
  {author} {\bibfnamefont {B.}~\bibnamefont {Gutkin}}, \ and\ \bibinfo {author}
  {\bibfnamefont {T.}~\bibnamefont {Guhr}},\ }\href@noop {} {\bibfield
  {journal} {\bibinfo  {journal} {ArXiv e-prints}\ } (\bibinfo {year}
  {2019})},\ \Eprint {http://arxiv.org/abs/1902.06265} {arXiv:1902.06265
  [cond-mat.stat-mech]} \BibitemShut {NoStop}%
\bibitem [{\citenamefont {Brouwer}\ and\ \citenamefont
  {Beenakker}(1996)}]{BnB}%
  \BibitemOpen
  \bibfield  {author} {\bibinfo {author} {\bibfnamefont {P.~W.}\ \bibnamefont
  {Brouwer}}\ and\ \bibinfo {author} {\bibfnamefont {C.~W.~J.}\ \bibnamefont
  {Beenakker}},\ }\href {\doibase 10.1063/1.531667} {\bibfield  {journal}
  {\bibinfo  {journal} {Journal of Mathematical Physics}\ }\textbf {\bibinfo
  {volume} {37}},\ \bibinfo {pages} {4904} (\bibinfo {year}
  {1996})}\BibitemShut {NoStop}%
\bibitem [{Note1()}]{Note1}%
  \BibitemOpen
  \bibinfo {note} {For finite $L$, only a discrete set of values of $s$ are
  allowed.}\BibitemShut {Stop}%
\bibitem [{Note2()}]{Note2}%
  \BibitemOpen
  \bibinfo {note} {See supplementary material at [url].}\BibitemShut {Stop}%
\bibitem [{\citenamefont {Sch\"utz}(2001)}]{SCHUTZ2001}%
  \BibitemOpen
  \bibfield  {author} {\bibinfo {author} {\bibfnamefont {G.}~\bibnamefont
  {Sch\"utz}},\ }in\ \href@noop {} {\emph {\bibinfo {booktitle} {Phase
  Transitions and Critical Phenomena}}},\ Vol.~\bibinfo {volume} {19},\
  \bibinfo {editor} {edited by\ \bibinfo {editor} {\bibfnamefont
  {C.}~\bibnamefont {Domb}}\ and\ \bibinfo {editor} {\bibfnamefont
  {J.}~\bibnamefont {Lebowitz}}}\ (\bibinfo  {publisher} {Academic Press},\
  \bibinfo {year} {2001})\BibitemShut {NoStop}%
\bibitem [{\citenamefont {Vanicat}\ \emph {et~al.}(2018)\citenamefont
  {Vanicat}, \citenamefont {Zadnik},\ and\ \citenamefont
  {Prosen}}]{ProsenIntTrot}%
  \BibitemOpen
  \bibfield  {author} {\bibinfo {author} {\bibfnamefont {M.}~\bibnamefont
  {Vanicat}}, \bibinfo {author} {\bibfnamefont {L.}~\bibnamefont {Zadnik}}, \
  and\ \bibinfo {author} {\bibfnamefont {T.~c.~v.}\ \bibnamefont {Prosen}},\
  }\href {\doibase 10.1103/PhysRevLett.121.030606} {\bibfield  {journal}
  {\bibinfo  {journal} {Phys. Rev. Lett.}\ }\textbf {\bibinfo {volume} {121}},\
  \bibinfo {pages} {030606} (\bibinfo {year} {2018})}\BibitemShut {NoStop}%
\bibitem [{\citenamefont {Takahashi}(2005)}]{TakahashiBook}%
  \BibitemOpen
  \bibfield  {author} {\bibinfo {author} {\bibfnamefont {M.}~\bibnamefont
  {Takahashi}},\ }\href@noop {} {\emph {\bibinfo {title} {Thermodynamics of
  one-dimensional solvable models}}}\ (\bibinfo  {publisher} {Cambridge
  University Press},\ \bibinfo {year} {2005})\BibitemShut {NoStop}%
\bibitem [{\citenamefont {Yang}\ and\ \citenamefont
  {Yang}(1969)}]{yang1969thermodynamics}%
  \BibitemOpen
  \bibfield  {author} {\bibinfo {author} {\bibfnamefont {C.-N.}\ \bibnamefont
  {Yang}}\ and\ \bibinfo {author} {\bibfnamefont {C.~P.}\ \bibnamefont
  {Yang}},\ }\href@noop {} {\bibfield  {journal} {\bibinfo  {journal} {Journal
  of Mathematical Physics}\ }\textbf {\bibinfo {volume} {10}},\ \bibinfo
  {pages} {1115} (\bibinfo {year} {1969})}\BibitemShut {NoStop}%
\bibitem [{\citenamefont {Suzuki}(2003)}]{suzuki2003quantum}%
  \BibitemOpen
  \bibfield  {author} {\bibinfo {author} {\bibfnamefont {M.}~\bibnamefont
  {Suzuki}},\ }\href@noop {} {\bibfield  {journal} {\bibinfo  {journal}
  {Physica A: Statistical Mechanics and its Applications}\ }\textbf {\bibinfo
  {volume} {321}},\ \bibinfo {pages} {334} (\bibinfo {year}
  {2003})}\BibitemShut {NoStop}%
\bibitem [{\citenamefont {Klümper}(2004)}]{klumper_2004}%
  \BibitemOpen
  \bibfield  {author} {\bibinfo {author} {\bibfnamefont {A.}~\bibnamefont
  {Klümper}},\ }\href {\doibase 10.1007/bfb0119598} {\bibfield  {journal}
  {\bibinfo  {journal} {Lecture Notes in Physics}\ ,\ \bibinfo {pages}
  {349–379}} (\bibinfo {year} {2004})}\BibitemShut {NoStop}%
\bibitem [{\citenamefont {Schlottmann}(1985)}]{schlottmann1985critical}%
  \BibitemOpen
  \bibfield  {author} {\bibinfo {author} {\bibfnamefont {P.}~\bibnamefont
  {Schlottmann}},\ }\href@noop {} {\bibfield  {journal} {\bibinfo  {journal}
  {Physical review letters}\ }\textbf {\bibinfo {volume} {54}},\ \bibinfo
  {pages} {2131} (\bibinfo {year} {1985})}\BibitemShut {NoStop}%
\bibitem [{\citenamefont {Takahashi}(1986)}]{takahashi1986quantum}%
  \BibitemOpen
  \bibfield  {author} {\bibinfo {author} {\bibfnamefont {M.}~\bibnamefont
  {Takahashi}},\ }\href@noop {} {\bibfield  {journal} {\bibinfo  {journal}
  {Progress of Theoretical Physics Supplement}\ }\textbf {\bibinfo {volume}
  {87}},\ \bibinfo {pages} {233} (\bibinfo {year} {1986})}\BibitemShut
  {NoStop}%
\bibitem [{\citenamefont {{{\v{S}}untajs}}\ \emph {et~al.}(2019)\citenamefont
  {{{\v{S}}untajs}}, \citenamefont {{Bon{\v{c}}a}}, \citenamefont {{Prosen}},\
  and\ \citenamefont {{Vidmar}}}]{Prosen2019mbl}%
  \BibitemOpen
  \bibfield  {author} {\bibinfo {author} {\bibfnamefont {J.}~\bibnamefont
  {{{\v{S}}untajs}}}, \bibinfo {author} {\bibfnamefont {J.}~\bibnamefont
  {{Bon{\v{c}}a}}}, \bibinfo {author} {\bibfnamefont {T.}~\bibnamefont
  {{Prosen}}}, \ and\ \bibinfo {author} {\bibfnamefont {L.}~\bibnamefont
  {{Vidmar}}},\ }\href@noop {} {\bibfield  {journal} {\bibinfo  {journal}
  {arXiv e-prints}\ ,\ \bibinfo {eid} {arXiv:1905.06345}} (\bibinfo {year}
  {2019})},\ \Eprint {http://arxiv.org/abs/1905.06345} {arXiv:1905.06345
  [cond-mat.str-el]} \BibitemShut {NoStop}%
\bibitem [{\citenamefont {Gaudin}(2014)}]{gaudin2014bethe}%
  \BibitemOpen
  \bibfield  {author} {\bibinfo {author} {\bibfnamefont {M.}~\bibnamefont
  {Gaudin}},\ }\href@noop {} {\emph {\bibinfo {title} {The Bethe
  Wavefunction}}}\ (\bibinfo  {publisher} {Cambridge University Press},\
  \bibinfo {year} {2014})\BibitemShut {NoStop}%
\bibitem [{\citenamefont {Franchini}()}]{franchini2017introduction}%
  \BibitemOpen
  \bibfield  {author} {\bibinfo {author} {\bibfnamefont {F.}~\bibnamefont
  {Franchini}},\ }\href@noop {} {\emph {\bibinfo {title} {An introduction to
  integrable techniques for one-dimensional quantum systems}}},\ Vol.\ \bibinfo
  {volume} {940}\ (\bibinfo  {publisher} {Springer})\BibitemShut {NoStop}%
\bibitem [{\citenamefont {Korepin}\ \emph {et~al.}(1997)\citenamefont
  {Korepin}, \citenamefont {Bogoliubov},\ and\ \citenamefont
  {Izergin}}]{korepin1997quantum}%
  \BibitemOpen
  \bibfield  {author} {\bibinfo {author} {\bibfnamefont {V.~E.}\ \bibnamefont
  {Korepin}}, \bibinfo {author} {\bibfnamefont {N.~M.}\ \bibnamefont
  {Bogoliubov}}, \ and\ \bibinfo {author} {\bibfnamefont {A.~G.}\ \bibnamefont
  {Izergin}},\ }\href@noop {} {\emph {\bibinfo {title} {Quantum inverse
  scattering method and correlation functions}}},\ Vol.~\bibinfo {volume} {3}\
  (\bibinfo  {publisher} {Cambridge university press},\ \bibinfo {year}
  {1997})\BibitemShut {NoStop}%
\bibitem [{\citenamefont {Faddeev}(1996)}]{faddeev1996algebraic}%
  \BibitemOpen
  \bibfield  {author} {\bibinfo {author} {\bibfnamefont {L.}~\bibnamefont
  {Faddeev}},\ }\href@noop {} {\bibfield  {journal} {\bibinfo  {journal} {arXiv
  preprint hep-th/9605187}\ } (\bibinfo {year} {1996})}\BibitemShut {NoStop}%
\bibitem [{\citenamefont {{Das}}\ \emph {et~al.}(2018)\citenamefont {{Das}},
  \citenamefont {{Damle}}, \citenamefont {{Dhar}}, \citenamefont {{Huse}},
  \citenamefont {{Kulkarni}}, \citenamefont {{Mendl}},\ and\ \citenamefont
  {{Spohn}}}]{nlfhHuse}%
  \BibitemOpen
  \bibfield  {author} {\bibinfo {author} {\bibfnamefont {A.}~\bibnamefont
  {{Das}}}, \bibinfo {author} {\bibfnamefont {K.}~\bibnamefont {{Damle}}},
  \bibinfo {author} {\bibfnamefont {A.}~\bibnamefont {{Dhar}}}, \bibinfo
  {author} {\bibfnamefont {D.~A.}\ \bibnamefont {{Huse}}}, \bibinfo {author}
  {\bibfnamefont {M.}~\bibnamefont {{Kulkarni}}}, \bibinfo {author}
  {\bibfnamefont {C.~B.}\ \bibnamefont {{Mendl}}}, \ and\ \bibinfo {author}
  {\bibfnamefont {H.}~\bibnamefont {{Spohn}}},\ }\href@noop {} {\bibfield
  {journal} {\bibinfo  {journal} {arXiv e-prints}\ ,\ \bibinfo {eid}
  {arXiv:1901.00024}} (\bibinfo {year} {2018})},\ \Eprint
  {http://arxiv.org/abs/1901.00024} {arXiv:1901.00024 [cond-mat.stat-mech]}
  \BibitemShut {NoStop}%
\bibitem [{\citenamefont {Gopalakrishnan}\ and\ \citenamefont
  {Vasseur}(2019)}]{PhysRevLett.122.127202}%
  \BibitemOpen
  \bibfield  {author} {\bibinfo {author} {\bibfnamefont {S.}~\bibnamefont
  {Gopalakrishnan}}\ and\ \bibinfo {author} {\bibfnamefont {R.}~\bibnamefont
  {Vasseur}},\ }\href {\doibase 10.1103/PhysRevLett.122.127202} {\bibfield
  {journal} {\bibinfo  {journal} {Phys. Rev. Lett.}\ }\textbf {\bibinfo
  {volume} {122}},\ \bibinfo {pages} {127202} (\bibinfo {year}
  {2019})}\BibitemShut {NoStop}%
\bibitem [{\citenamefont {{De Nardis}}\ \emph {et~al.}(2019)\citenamefont {{De
  Nardis}}, \citenamefont {{Medenjak}}, \citenamefont {{Karrasch}},\ and\
  \citenamefont {{Ilievski}}}]{anomalousDN}%
  \BibitemOpen
  \bibfield  {author} {\bibinfo {author} {\bibfnamefont {J.}~\bibnamefont {{De
  Nardis}}}, \bibinfo {author} {\bibfnamefont {M.}~\bibnamefont {{Medenjak}}},
  \bibinfo {author} {\bibfnamefont {C.}~\bibnamefont {{Karrasch}}}, \ and\
  \bibinfo {author} {\bibfnamefont {E.}~\bibnamefont {{Ilievski}}},\
  }\href@noop {} {\bibfield  {journal} {\bibinfo  {journal} {arXiv e-prints}\
  ,\ \bibinfo {eid} {arXiv:1903.07598}} (\bibinfo {year} {2019})},\ \Eprint
  {http://arxiv.org/abs/1903.07598} {arXiv:1903.07598 [cond-mat.stat-mech]}
  \BibitemShut {NoStop}%
\end{thebibliography}%

\bigskip

\onecolumngrid
\newpage



\section*{Supplementary material}

# Supplemental Material for "Spectral statistics and many-body quantum chaos with conserved charge"

Aaron J. Friedman,[1,2] Amos Chan,[1] Andrea De Luca,[1,3] and J. T. Chalker[1]

[1]*Rudolf Peierls Centre for Theoretical Physics, Clarendon Laboratory, University of Oxford, Oxford, OX1 3PU, UK*
[2]*Department of Physics and Astronomy, University of California, Irvine, CA 92697, USA*
[3]*Laboratoire de Physique Théorique et Modélisation (CNRS UMR 8089),
Université de Cergy-Pontoise, F-95302 Cergy-Pontoise, France*
(Dated: October 16, 2019)

## CONTENTS

## A. MAPPING BETWEEN $K(t)$ AND THE TROTTERIZED XXX HEISENBERG CHAIN

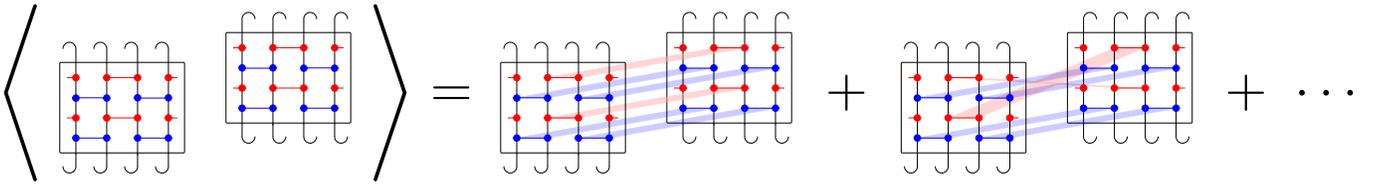

FIG. S1. A schematic illustration of the diagrammatic representation of $K(t,s)$ according to [1].

According to [1], the evaluation of the ensemble average in the definition of $K(t,s)$ (in Eq. (2) of the main text) amounts to generating all diagrams by pairing unitaries with their complex conjugates at each bond, as schematically illustrated in Fig. S1. In the large-$q$ limit, the leading contributions consist of $t$ diagrams, each of which has an identical 'cyclical' pairing at *all* sites as illustrated in Fig. S2a (for a rigorous proof, see [1]).

These $t$ diagrams are 'Gaussian', in the sense that they can be evaluated using the Wick contraction of a random unitary and its conjugate (Fig. S2b),

$$\overline{U^{k,\alpha;l,\alpha}_{i,\alpha;j,\alpha} U^{*\,k',\alpha;l',\alpha}_{i',\alpha;j',\alpha}} = \frac{1}{q^2} \delta_{i,i'} \delta_{j,j'} \delta_{k,k'} \delta_{l,l'}$$
$$\overline{U^{k,\alpha;l,\beta}_{i,\alpha;j,\beta} U^{*\,k',\alpha;l',\beta}_{i',\alpha;j',\beta}} = \frac{1}{2q^2} \delta_{i,i'} \delta_{j,j'} \delta_{k,k'} \delta_{l,l'} \qquad \text{if } \alpha \neq \beta \qquad , \tag{S1}$$

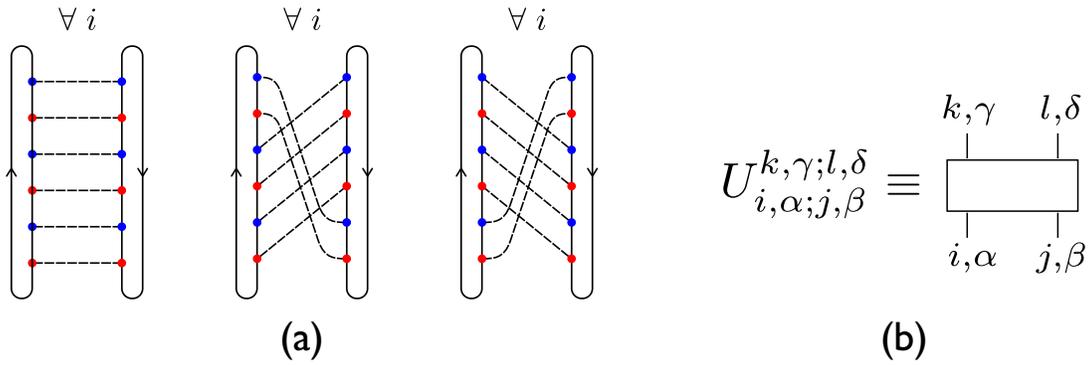

FIG. S2. (a): The three leading diagrams of $K(t,s)$ at $t=3$ in the large-$q$ limit [1]. For each diagram, every site $i$ takes the *same* configuration. (b): The diagrammatical representation of $U^{k,\gamma;l\delta}_{i,\alpha;j,\beta}$, where the Roman [Greek] indices correspond to color [spin] degrees of freedom.

where Roman [Greek] indices correspond to color [spin] degrees of freedom. Using (S1), we translate each of these $t$ diagrams into an algebraic term by summing over the color and spin degrees of freedom. For simplicity, consider the leftmost diagram in Fig. S2a. The sum over the color degrees of freedom will cancel out all factors of $q$ that appear in (S1), as shown in [1]. This leaves a sum over the spin degrees of freedom, subject to the constraint $S^z = \sum_{j=1}^{L} \hat{\sigma}_j^z = sL$. Observe that the choice of spin degrees of freedom in $\hat{W}$ fixes those in $\hat{W}^\dagger$ due to (S1). Consequently, the sum over spins-1/2 can be computed by finding all possible ways to assign spins in the diagrammatic representation of $\mathrm{Tr}_s[\hat{W}]$ – as illustrated in Fig. S3 – such that ($i$) the global magnetization satisfies $S^z = Ls$ along any horizontal fixed-time slice of the diagram, and ($ii$) the local magnetization $S^z_{j,j+1} \equiv \hat{\sigma}^z_j + \hat{\sigma}^z_{j+1}$ is preserved across each two-site gate. This sum over spins can be reproduced by $\mathrm{Tr}_s[\hat{M}^t]$, as defined in Eq. (3) of the main text, where the factor of $\frac{1}{2}$ in $\hat{\mathbb{1}}_{j,j'}$ originates from the $\frac{1}{2q^2}$ term in (S1), while the $q^2$ factor is cancelled by averaging over the color degrees of freedom.

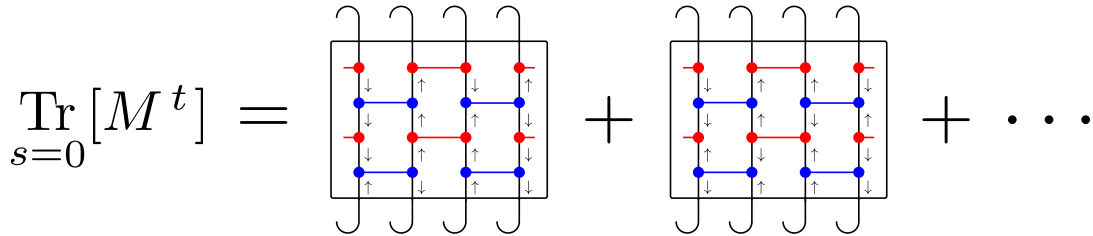

FIG. S3. The sum over spins can be computed by finding all possible configurations of the spins-1/2 allowed in the diagrammatic representation of $\mathrm{Tr}_s[\hat{W}]$, such that the global magnetization satisfies $S^z = sL$ ($s=0$ in this figure), and the local magnetization is conserved by each gate.

Finally, note that any two of the $t$ contracted diagrams in Fig. S2a can be related to one another by a "rotation" of the arrowed loop on the right, i.e. the diagrams are topologically equivalent. This loop represents a trace, and the aforementioned invariance is equivalent to the cyclic invariance of the trace. Consequently, the corresponding algebraic terms for all $t$ leading diagrams are *identical*, giving rise to the factor of $t$ and reproducing Eq. (3) of the main text in full.

## B. NUMERICAL METHODS

For $L = 10, 12, 14$, we find the eigenvalues of $\hat{W}$, given by Eq. (11) of the main text, using exact diagonalization, and compute $\mathrm{Tr}[\hat{W}(t)] = \sum_n e^{it\theta_n}$ directly to obtain $K(t)$, which is then averaged. For $L = 16$ and $18$, we sample the trace, computing an approximate version of $\mathrm{Tr}[\hat{W}(t)]$ via successive multiplication of $\hat{W}$ onto a number of 'sample states', which are superpositions of the basis states, with complex coefficients drawn from a Gaussian distribution with zero mean and unit variance. One can either sample the trace within a given realization of spatial disorder, or

sample both disorder realizations and the trace concurrently. In either case, we expect the *total* number of samples required to reduce noise to a given degree to be the same for either method.

For the 12870 basis states comprising the trace for $L = 16$, we find that $\lesssim 100$ samples per disorder realization is sufficient to reproduce the correct behavior for $K(t)$, with $\gtrsim 10^4$ realizations of disorder. For $L = 18$, we use concurrent sampling of disorder and the trace; owing to the larger Hilbert space dimension, to obtain results with noise comparable to the $L = 16$ data would likely close to $10^7$ combined samples and realizations. The data shown in the main text correspond to $4 \times 10^5$ combined samples and realizations of disorder. As these data are noisier than those for smaller system sizes, we also average this data in windows of roughly 25 time steps to reduce noise further; we plot the average within each window, accompanied by error bars corresponding to the combination of disorder averaging and this temporal smoothing.

### C. GENERAL FORMULATION OF INHOMOGENEOUS ALGEBRAIC BETHE ANSATZ

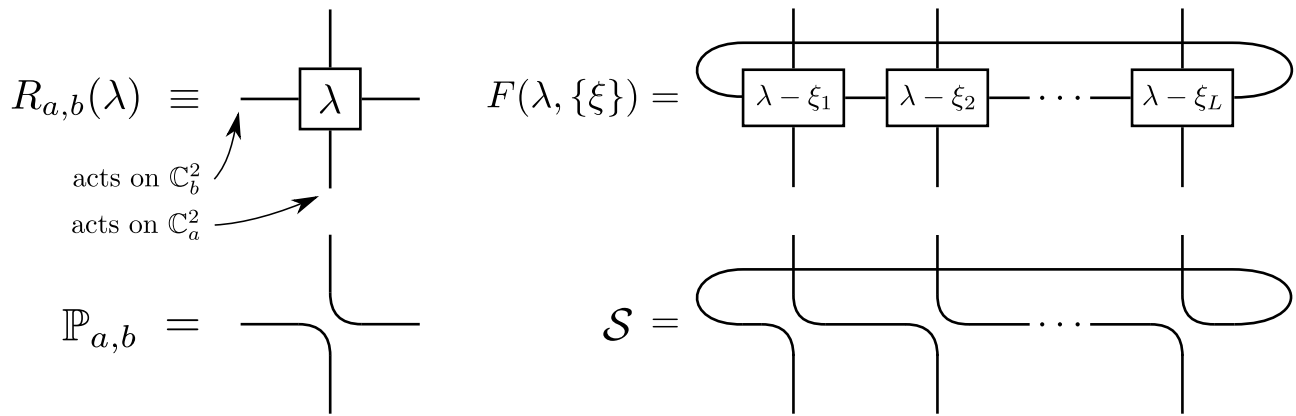

FIG. S4. Diagrammatic representations of $\hat{\mathcal{R}}_{a,b}(\lambda-\mu)$, $\hat{F}(\lambda,\{\xi\})$, $\hat{\mathbb{P}}_{a,b}$ and $\hat{\mathcal{S}}$. Note that in the current convention, $\hat{\mathcal{R}}_{a,b}(\lambda-\mu)$ acts in the top right direction.

In Eq. (13) of the main text, we introduced the standard $SU(2)$ $R$-matrix (Fig. S4). The exchange operator $\hat{\mathbb{P}}_{a,b}$ is defined by $\hat{\mathbb{P}}_{a,b}|s_a, s_b\rangle = |s_b, s_a\rangle$. The fundamental property is that the $R$-matrix satisfies the Yang-Baxter identity. For clarity, indices of the form $i, j, k, \ldots$ will indicate sites in the real spin lattice, while indexes of the form $a, b, c, \ldots$ will indicate the auxiliary spaces, all of which are spins-1/2. The Yang-Baxter identity can be written as an operator identity in $\mathbb{C}_a^2 \otimes \mathbb{C}_b^2 \otimes \mathbb{C}_i^2$, and we have

$$\hat{\mathcal{R}}_{a,b}(\lambda-\mu)\hat{\mathcal{R}}_{i,a}(\lambda)\hat{\mathcal{R}}_{i,b}(\mu) = \hat{\mathcal{R}}_{i,b}(\mu)\hat{\mathcal{R}}_{i,a}(\lambda)\hat{\mathcal{R}}_{a,b}(\lambda-\mu) \tag{S2}$$

One can then introduce the $T$-matrix as

$$\hat{\mathcal{T}}_{L,a}(\lambda,\{\xi\}) = \hat{\mathcal{R}}_{1,a}(\lambda-\xi_1)\hat{\mathcal{R}}_{2,a}(\lambda-\xi_2)\ldots\hat{\mathcal{R}}_{L,a}(\lambda-\xi_L) \tag{S3}$$

Note that with respect to Eq. (14) in the main text, we write explicitly its dependence on the set of inhomogeneities $\xi$'s. The crucial fact is that employing (S2) many times, one can prove the so-called *RTT* relation

$$\hat{\mathcal{R}}_{a,b}(\lambda-\mu)\hat{\mathcal{T}}_{L,a}(\lambda,\{\xi\})\hat{\mathcal{T}}_{L,b}(\mu,\{\xi\}) = \hat{\mathcal{T}}_{L,b}(\mu,\{\xi\})\hat{\mathcal{T}}_{L,a}(\lambda,\{\xi\})\hat{\mathcal{R}}_{a,b}(\lambda-\mu) \tag{S4}$$

From this relation it follows that if we set $F(\lambda,\{\xi\})$ as the trace over the auxiliary space, i.e. $F(\lambda,\{\xi\}) = \text{Tr}_a[\hat{\mathcal{T}}_{L,a}(\lambda,\{\xi\})]$, as illustrated in Fig. S4, we have that

$$[F(\lambda,\{\xi\}), F(\lambda',\{\xi\})] = 0, \qquad \lambda, \lambda' \in \mathbb{C}.$$

Repeating the procedure of Eq. (15) of the main text, but keeping explicit the dependence on $\xi$'s

$$\hat{\mathcal{T}}_{L,a}(\lambda,\{\xi\}) = \begin{pmatrix} A(\lambda,\{\xi\}) & B(\lambda,\{\xi\}) \\ C(\lambda,\{\xi\}) & D(\lambda,\{\xi\}) \end{pmatrix}. \tag{S5}$$

With this notation it is clear that $F(\lambda, \{\xi\}) = A(\lambda, \{\xi\}) + D(\lambda, \{\xi\})$ and moreover (S4) can be interpreted as a set of commutation relations for the operators $A, B, C, D$. Those relevant for us are

$$[B(\lambda, \{\xi\}), B(\mu, \{\xi\})] = 0 \tag{S6}$$

$$A(\lambda, \{\xi\})B(\mu, \{\xi\}) = f_+(\lambda - \mu)B(\mu, \{\xi\})A(\lambda, \{\xi\}) + g_+(\lambda - \mu)B(\lambda, \{\xi\})A(\mu, \{\xi\}) \tag{S7}$$

$$D(\lambda, \{\xi\})B(\mu, \{\xi\}) = f_-(\lambda - \mu)B(\mu, \{\xi\})D(\lambda, \{\xi\}) + g_-(\lambda - \mu)B(\lambda, \{\xi\})D(\mu, \{\xi\}) \tag{S8}$$

where the functions

$$f_\pm(\lambda) = \frac{\lambda \mp 2\imath}{\lambda}, \qquad g_\pm(\lambda) = \pm\frac{2\imath}{\lambda} \tag{S9}$$

are independent of the inhomogeneities $\{\xi\}$. We fix the vacuum state $|\Omega\rangle = |\uparrow\uparrow\ldots\uparrow\rangle$. This is an eigenstate of $F(\lambda, \{\xi\})$, since

$$A(\lambda, \{\xi\})|\Omega\rangle = a(\lambda, \{\xi\})|\Omega\rangle; \qquad D(\lambda, \{\xi\})|\Omega\rangle = d(\lambda, \{\xi\})|\Omega\rangle; \qquad C(\lambda, \{\xi\})|\Omega\rangle = 0. \tag{S10}$$

And with our current normalizations, we have

$$a(\lambda, \{\xi\}) = 1, \tag{S11}$$

$$d(\lambda, \{\xi\}) = \prod_i \left(\frac{\lambda - \xi_i}{\lambda - \xi_i + 2\imath}\right); \tag{S12}$$

Then one can check [2] that the operators $B(\lambda, \{\xi\})$ can be used to build eigenstates of $F(\lambda, \{\xi\})$

$$F(\lambda, \{\xi\})|\{\lambda\}\rangle = \Lambda(\lambda, \{\lambda\}, \{\xi\})|\{\lambda\}\rangle, \tag{S13}$$

$$|\{\lambda\}\rangle \equiv B(\lambda_1, \{\xi\})\ldots B(\lambda_N, \{\xi\})|\Omega\rangle \tag{S14}$$

if the Bethe Ansatz equations are satisfied

$$\prod_{b \neq a} \frac{f_-(\lambda_a - \lambda_b)}{f_+(\lambda_a - \lambda_b)} = \frac{a(\lambda_a, \{\xi\})}{d(\lambda_a, \{\xi\})} \tag{S15}$$

where $a, b = 1, \ldots, N$. The eigenvalue $\Lambda(\lambda, \{\lambda\}, \{\xi\})$ is given by

$$\Lambda(\lambda, \{\lambda\}, \{\xi\}) = a(\lambda, \{\xi\})\prod_a f_+(\lambda - \lambda_a) + d(\lambda, \{\xi\})\prod_a f_-(\lambda - \lambda_a) \tag{S16}$$

Note that each $B(\lambda, \{\xi\})$ decreases by 1 the total magnetization along $z$:

$$[B(\lambda), \hat{S}^z] = B(\lambda) \implies \hat{S}^z|\{\lambda\}\rangle = (L/2 - N)|\{\lambda\}\rangle. \tag{S17}$$

### D. TROTTERIZED XXX

We now fix the values of the inhomogeneities according to

$$\xi_j = \imath + (-1)^j \delta \tag{S18}$$

and for clarity, we refer to the transfer matrix for the Trotterized problem as $F_\delta(\lambda) \equiv F(\lambda, \{\xi\})$ with the $\xi_j$'s given by (S18). We also assume $L$ to be an even integer. Then we define the evolution operator as

$$\hat{M}_\delta = F_\delta(\imath - \delta)^{-1} F_\delta(\imath + \delta). \tag{S19}$$

Inspecting (S3), we see that it involves the matrices

$$\hat{\mathcal{R}}_{a,b}(0) = \mathbb{P}_{a,b}, \quad \hat{\mathcal{R}}_{a,b}(\pm\delta) \equiv \hat{\mathcal{R}}^\pm_{a,b}. \tag{S20}$$

We also introduce the shift operator by

$$\mathcal{S}|s_1, s_2, \ldots, s_L\rangle = |s_2, s_3, \ldots s_L, s_1\rangle, \qquad \mathcal{S}^\dagger \mathcal{S} = \mathbf{1}, \tag{S21}$$

which can be rewritten as a trace over the auxiliary space

$$\mathcal{S} = \text{Tr}_a[\mathbb{P}_{1,a}\mathbb{P}_{2,a}\ldots\mathbb{P}_{L,a}] \tag{S22}$$

Then we can write

$$\mathcal{S}^\dagger F_\delta(\imath + \delta) = \check{R}^+_{1,2}\check{R}^+_{3,4}\ldots\check{R}^+_{L-1,L} , \tag{S23a}$$

$$\mathcal{S}^\dagger F_\delta(\imath - \delta) = \check{R}^-_{L,1}\check{R}^-_{2,3}\ldots\check{R}^-_{L-2,L-1} , \tag{S23b}$$

where we introduced

$$\check{R}^\pm_{ab} = \hat{\mathbb{P}}_{ab}\hat{\mathcal{R}}^\pm_{ab} \tag{S24}$$

Note that (S23) also imply that the square of the shift operator belongs to the set of commuting operators

$$\mathcal{S}^2 = F_\delta(\imath + \delta)F_\delta(\imath - \delta) , \tag{S25}$$

as expected from the two-site translation invariance of the model. Moreover, from (S23b) we immediately read off the expression for $F(\imath/2 - \delta/2)^{-1}$ as

$$F_\delta(\imath - \delta)^{-1} = (\check{R}^-_{L-2,L-1})^{-1}(\check{R}^-_{L-4,L-3})^{-1}\ldots(\check{R}^-_{2,3})^{-1}(\check{R}^-_{L,1})^{-1}\mathcal{S}^\dagger \tag{S26}$$

Furthermore, from Eq. (13) of the main text, we see that

$$(\hat{\mathcal{R}}^\pm_{a,b})^{-1} = \hat{\mathcal{R}}^\mp_{a,b} . \tag{S27}$$

Therefore, we conclude that

$$\hat{M}_\delta = \check{R}^+_{L-2,L-1}\check{R}^+_{L-4,L-3}\ldots\check{R}^+_{L,1}\check{R}^+_{1,2}\check{R}^+_{3,4}\ldots\check{R}^+_{L-1,L} . \tag{S28}$$

Thus, inserting (S18) into the general Bethe Ansatz equation (S15), one obtains

$$\left(\frac{\lambda_a + (\delta + \imath)}{\lambda_a + (\delta - \imath)}\right)^{L/2}\left(\frac{\lambda_a - (\delta - \imath)}{\lambda_a - (\delta + \imath)}\right)^{L/2} = \prod_{\substack{b=1 \\ b\neq a}}^N \left(\frac{\lambda_a - \lambda_b + 2\imath}{\lambda_a - \lambda_b - 2\imath}\right) . \tag{S29}$$

The eigenvalue of $\hat{M}_\delta$ corresponding to the state $|\{\lambda\}\rangle$ is then given by

$$\hat{M}_\delta|\{\lambda\}\rangle = \prod_a \frac{f_+(\imath + \delta - \lambda_a)}{f_+(\imath - \delta - \lambda_a)}|\{\lambda\}\rangle , \tag{S30}$$

where we used the relation $d(\frac{\imath\pm\delta}{2}) = 0$. We now observe that for $\delta = \imath$, $\check{R}^+ = \frac{1}{2}(\hat{\mathbb{1}} + \hat{\mathbb{P}})$ and thus

$$\hat{M} = \lim_{\delta\to\imath}\hat{M}_\delta . \tag{S31}$$

In practice, while varying the parameter $\delta \in [0, \imath]$ we interpolate between the standard XXX spin chain and the model in Eq. (3) of the main text. In this limit, the Bethe Ansatz equation (S29) reproduces Eq. (18) in the main text. The eigenvalues in (S30) can be written as

$$\hat{M}B(\lambda_1)\ldots B(\lambda_N)|\Omega\rangle = e^{\sum_a -\varepsilon(\lambda_a)}|\Omega\rangle , \qquad \varepsilon(\lambda) = -\log\frac{\lambda^2}{\lambda^2 + 4} \tag{S32}$$

where $\varepsilon(\lambda)$ plays the role of an effective magnon energy. Using (S25), we obtain also the spectrum of the shift operator

$$\mathcal{S}^2|\{\lambda\}\rangle = e^{2\imath\sum_a k(\lambda_a)}|\{\lambda\}\rangle , \qquad e^{2\imath k(\lambda)} = \left(\frac{\lambda + 2\imath}{\lambda - 2\imath}\right) \tag{S33}$$

which coincides with the relation $k(\lambda) = \text{arccot}(\lambda/2) \in [-\pi/2, \pi/2]$ given in the main text. In terms of it the dispersion relation in (S32) is rewritten as $\varepsilon(\lambda) = 2\ln\cos k(\lambda)$ as stated in the main text.

## E.  STRING HYPOTHESIS

To recover the other solutions to the Bethe Ansatz Equations (BAE), i.e. Eq. (18) of the main text, we apply the string hypothesis [3]. We assume that the standard strings can be used, since the scattering matrix (right-hand side of the BAE) is unchanged compared to the standard XXX spin chain. Therefore, we assume that the roots of BAE are arranged as

$$\lambda_\alpha^{(n,m)} = \lambda_\alpha^{(n)} + \imath(2m - n + 1), \qquad m = 0, \ldots, n-1 \tag{S34}$$

where $n$ is the number of roots contained in the $n$-string. The index $\alpha$ labels the strings of the same type and $\lambda_\alpha^{(n)} \in \mathbb{R}$ is the string center. Deviations with respect to (S34) are expected to be exponentially small in the system size $L$. Here, we assume that the string hypothesis is valid at least for generic values of the parameter $\delta$, that we keep as a regularizer that we send to $\imath$.

Inserting the string Ansatz into the BAE (Eq. (18) main text), we recover the string form of the Bethe Ansatz equation which involves only the string centers. The derivation is completely analogous to the one holding for the XXX spin chain so we skip the details and refer the reader to standard references (see for instance chapter 8 in [4]). One then obtains the logarithmic form of the Bethe Ansatz equations for the string centers

$$L\,\theta_n^{(\delta)}(\lambda_\alpha^{(n)}) = 2\pi I_\alpha^{(n)} + \sum_{(m,\beta) \neq (n,\alpha)} \Theta_{n,m}(\lambda_\alpha^n - \lambda_\beta^m). \tag{S35}$$

This equation is almost identical to the one holding for the standard XXX spin chain, upon replacing $\theta(\lambda) \to \theta^{(\delta)}(\lambda)$, where for an arbitrary function $g(\lambda)$ we defined

$$g^{(\delta)}(\lambda) = \frac{1}{2}\left(g(\lambda + \delta) + g(\lambda - \delta)\right) \tag{S36}$$

and we set

$$\theta_n(\lambda) = 2\arctan(\lambda/n). \tag{S37}$$

We then take the thermodynamic limit $L \to \infty$ of (S35) in a standard way, by introducing the density of roots and holes for each string length $n$, respectively $\rho_n(\lambda)$ and $\rho_n^h(\lambda)$. We refer the reader to the original references for their precise definitions. For our purposes, the ensemble of root densities characterize each eigenstate in thermodynamic limit, as the density of the string centers

$$\rho_n(\lambda) = \frac{1}{L}\sum_\alpha \delta(\lambda - \lambda_\alpha^{(n)}), \qquad n = 1, \ldots, \infty. \tag{S38}$$

From this it follows that the eigenvalues of $\hat{M}$ and the total magnetization $\hat{S}^z$ converge to

$$\langle\{\lambda\}|\hat{M}|\{\lambda\}\rangle \stackrel{L\to\infty}{=} \exp\left[L\sum_{n=1}^\infty \varepsilon_n(\lambda)\rho_n(\lambda)\right] \tag{S39}$$

$$\langle\{\lambda\}|S^z|\{\lambda\}\rangle \stackrel{L\to\infty}{=} \frac{L}{2} - L\sum_{n=1}^\infty n\rho_n(\lambda), \tag{S40}$$

where we defined the effective energy of $n$-string magnon

$$\epsilon_n(\lambda) \equiv \sum_{m=0}^{n-1} \epsilon\bigl(\lambda + \imath(2m - n + 1)\bigr) = \ln\left(\frac{\lambda^2 + (n-1)^2}{\lambda^2 + (n+1)^2}\right). \tag{S41}$$

The Bethe Ansatz equations (S35) translate into a functional relation between the density of roots and the density of holes

$$a_n^{(\delta)}(\lambda) = \rho_n(\lambda) + \rho_n^h(\lambda) + \sum_{m=1}^\infty (T_{nm} * \rho_m)(\lambda), \qquad n = 1, \ldots, \infty \tag{S42}$$

where we denote the convolution as

$$(g * h)(\lambda) = \int d\lambda' g(\lambda - \lambda') h(\lambda')$$

and we defined the quantities

$$a_n(\lambda) = \frac{1}{2\pi} \frac{d}{d\lambda} \theta(\lambda/n) = \frac{1}{\pi} \frac{n}{\lambda^2 + n^2} \tag{S43}$$

$$T_{nm}(\lambda) = a_{|n-m|}(\lambda)(1 - \delta_{n,m}) + a_{n+m}(\lambda) + 2 \sum_{\ell=|n-m|+1}^{n+m-1} a_\ell(\lambda) . \tag{S44}$$

### F. THERMODYNAMIC BETHE ANSATZ (TBA)

We now consider the evaluation of the function $\phi(t, s)$ defined in Eq. (7) of the main text, which we repeat for convenience

$$\phi(t, s) = -\lim_{L \to \infty} L^{-1} \ln \operatorname*{Tr}_s(\hat{M}^t) \tag{S45}$$

To account for the projection onto the subspace $S = sL$, we make use of the Legendre transform, namely

$$\phi(t, s) = \max_h [\psi(t, h) - 2hs] , \qquad \psi(t, h) = -\lim_{L \to \infty} L^{-1} \log \operatorname{Tr}(\hat{M}^t e^{2hS^z}) \tag{S46}$$

where the trace in (S46) is unrestricted. We explain here how $\psi(t, h)$ can be exactly computed within the formalism of the TBA. The standard procedure is to rewrite the trace as a functional integral over the root/holes densities

$$\operatorname{Tr}[M^t e^{-2hS^z}] = \sum_{\{\lambda\}} \langle \{\lambda\}| M^t e^{2hS^z} |\{\lambda\}\rangle = \int \mathcal{D}\rho_n \mathcal{D}\rho_n^h \, \delta(\text{BAE}) e^{-L\mathcal{F}[\rho, \rho^h]} \tag{S47}$$

where the $\delta(\text{BAE})$ enforces the relation (S42). The functional $\mathcal{F}(\rho, \rho^h)$ takes the form

$$\mathcal{F}[\rho, \rho^h] = -h + \sum_n \int dx \rho_n(\lambda)(\epsilon_n(\lambda) + 2hn) + S_{YY}[\rho, \rho^h] \tag{S48}$$

$$S_{YY}[\rho, \rho^h] = -\rho_n(\lambda) \ln\left(1 + \frac{\rho_n^h(\lambda)}{\rho_n(\lambda)}\right) - \rho_n^h(\lambda) \ln\left(1 + \frac{\rho_n(\lambda)}{\rho_n^h(\lambda)}\right) \tag{S49}$$

where the Yang-Yang entropy $S_{YY}[\rho, \rho^h]$ [5] accounts for the exponentially large number of states described by the same root densities. At large $L$, the minimization of $\mathcal{F}[\rho, \rho^h]$ leads to the equations for $\eta_n(\lambda) = \rho_n^h(\lambda)/\rho_n(\lambda)$

$$\ln \eta_n(\lambda) = t\epsilon_n(\lambda) + 2hn + \sum_{m=1}^{\infty} T_{nm} * \ln(1 + \eta_m^{-1}(\lambda)) . \tag{S50}$$

Inserting this solution in the expression for $\mathcal{F}[\rho, \rho^h]$ we get its value at the minimum, namely

$$\psi(t, h) = -h - \sum_n \int dx a_n^{(\delta)}(\lambda) \ln\left(1 + \eta_n^{-1}(\lambda)\right) \tag{S51}$$

This equation holds generically for any $\delta$. It can be further simplified for $\delta \to \imath$. Indeed, writing explicitly the case $n = 1$ of (S50):

$$\ln(1 + \eta_1(\lambda)) = tg_1^{(\delta)}(\lambda) + 2h + \sum_{m=1}^{\infty} (a_{m-1} + a_{m+1}) * \ln(1 + \eta_m^{-1}(\lambda)) . \tag{S52}$$

We moreover note that

$$\frac{a_{m-1}(\lambda) + a_{m+1}(\lambda)}{2} = a_m^{(\imath)}(\lambda) \tag{S53}$$

we obtain

$$\psi(t,h) \equiv \lim_{\lambda \to 0} \frac{t\epsilon_1(\lambda) - \ln \eta_1(\lambda)}{2} \tag{S54}$$

Note that the limiting procedure is required just because both $g_1^{(i)}(\lambda)$ and $\ln \eta_1(\lambda)$ are singular for $\lambda \to 0$ but their difference is not.

## G. LARGE-TIME EXPANSION

We consider the asymptotic behavior of $\psi(t,h)$ for large $t$ and fixed $h$. Since $\psi(t,-h) = \psi(t,h)$, we take $h > 0$. The case $h = 0$ requires a different treatment. The functions $\eta_n(\lambda)$ admit the expansion:

$$\ln \eta_n(\lambda) = t\epsilon_n(\lambda) + 2nh + \sum_{k=0}^{\infty} t^{-k/2} q_{n,k}(\lambda/\sqrt{t}) \tag{S55}$$

The following results are easily verified

$$\int_{-\infty}^{\infty} dx\, a_n(\lambda) = 1 \,, \qquad \lim_{t \to \infty} t a_n(\sqrt{t}z) = \frac{n}{\pi z^2} \,. \tag{S56}$$

We have therefore

$$\sqrt{t}\, T_{nm}(z\sqrt{t}) \stackrel{t \gg 1}{=} \tau_{n,m} \delta(z) + \frac{2nm}{\pi \sqrt{t} z^2} + O(t^{-3/2}) \,, \tag{S57}$$

$$\lim_{t \to \infty} \ln \eta_n(\sqrt{t}z) = \frac{4n}{z^2} + 2nh + q_{n,0}(z) \,, \tag{S58}$$

where

$$\tau_{n,m} = \int dx\, T_{n,m}(\lambda) = 2\min(n,m) - \delta_{n,m}$$

Indeed, inserting (S55) inside (S50), we have at the leading order an equation for $q_{n,0}(z)$

$$q_{n,0}(z) = \sum_{m=1}^{\infty} \tau_{n,m} \ln\left(1 + e^{-\frac{4m}{z^2} - 2mh - e_m^{(0)}(z)}\right) \tag{S59}$$

These equations can be mapped onto those describing the infinite temperature behavior of the XXX model and admit an exact solution in the form

$$q_{n,0}(z) = -\frac{4m}{z^2} - 2mh + \log\left(\frac{\sinh(n(h + 2z^{-2}))\sinh((n+2)(h + 2z^{-2}))}{\sinh(h + 2z^{-2})^2}\right) \,. \tag{S60}$$

In particular, this implies $q_{n,0}(z \to 0) = 0$. At the next order, we have an equation for $q_{n,1}(z)$

$$q_{n,1}(z) = -\sum_{m=1}^{\infty} \frac{\tau_{nm} q_{m,1}(z)}{1 + e^{\frac{4m}{z^2} + 2mh + q_{m,0}(z)}} + \sum_{m=1}^{\infty} \int dz'\, \frac{2nm}{\pi(z - z')^2} \ln(1 + e^{-\frac{4m}{z'^2} - 2mh - q_{m,0}(z')}) \tag{S61}$$

So using (S54), we need only the case $n = 1$. Changing variable $u = h + 2/z^2$, we obtain

$$\psi(t,h) \stackrel{t \gg 1}{=} -h - \frac{q_{1,1}(0)}{2\sqrt{t}} = -h - \sum_{m=1}^{\infty} \int_h^{\infty} dh\, \frac{m \log\left(\frac{\sinh((m+1)u)^2}{\sinh(mu)\sinh((m+2)u)}\right)}{\sqrt{2\pi^2 t(u - h)}} \tag{S62}$$

Remarkably, exchanging the sum and the integral allows for an explicit solution: setting $g_m = \log(\sinh(mu))$ and using

$$\sum_{m=1}^{M} m(2g_{m+1} - g_m - g_{m+1}) = (M+1)g_{M+1} - Mg_M - g_1 \tag{S63}$$

one recovers

$$\psi(t,h) = -h - \frac{\text{Li}_{3/2}\left(e^{-2h}\right)}{2\sqrt{\pi t}} + o(t^{-1/2}) \,. \tag{S64}$$

The Legendre transform is obtained by observing that at large $t$, the maximum in (S46) is at $h \sim 0$. So one expands

$$\text{Li}_{3/2}\left(e^{-2h}\right) = 2\sqrt{2\pi h} + \zeta(3/2) + O(h) \,, \tag{S65}$$

and find that (S46) is maximal at $h^* \simeq 1/(2(1+2s)^2 t)$. Combining this with (S46) recovers Eq. (8) of the main text.

## H. ALTERNATIVE APPROACH: QUANTUM TRANSFER MATRIX (QTM)

The formulation presented in the previous section allows the calculation of the large-time expansion. However, the calculation at arbitrary values of $t$ and $h$ of $\psi(t,h)$ requires the solution of (S50), which is an infinite set of coupled integral equations. One has to resort to truncation $n$ up to when convergence is reached. A numerically much more efficient procedure is based on the *quantum transfer matrix* approach [6]. We first note the relation

$$F_\delta(\imath - \delta)^{-1} = \lim_{x \to -\imath - \delta} d(x)^{-1} F_\delta(x) \tag{S66}$$

Then from (S19), we can write

$$\text{Tr}[M_\delta^t e^{2hS^z}] = \lim_{x \to -\imath - \delta} d(x)^{-t} \text{Tr}[\underbrace{F_\delta(x) F_\delta(\imath + \delta) \ldots F_\delta(x) F_\delta(\imath + \delta)}_{t \text{ times}} e^{2hS^z}] = \text{Tr}[\underbrace{\tilde{F}_1 \tilde{F}_2 \ldots \tilde{F}_1 \tilde{F}_2}_{L \text{ times}}] \tag{S67}$$

where $\tilde{F}_{1,2}$ are dual transfer matrices defined on $2t$ sites, obtained via a 90° space-time rotation, which exchanges space and time, as clarified in Fig. S5. The advantage of this transformation is that in the limit $L \to \infty$, the last

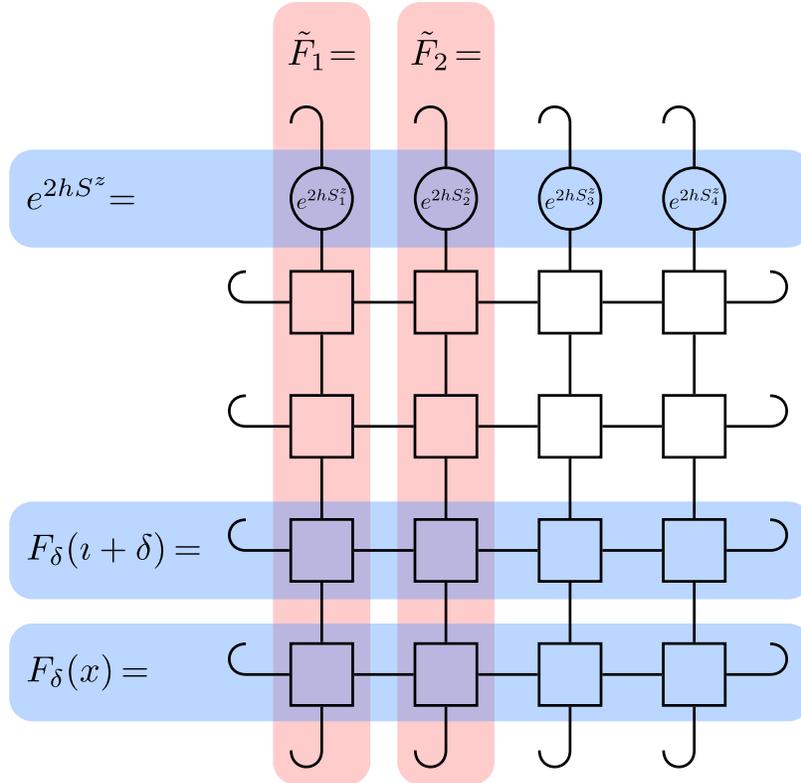

FIG. S5. Diagrammatic representation of $\text{Tr}[M_\delta^t e^{2hS^z}]$. Blue (red) boxes group matrices that act in the time (space) direction. The curly legs at the boundary of the diagram represent the periodic boundary condition.

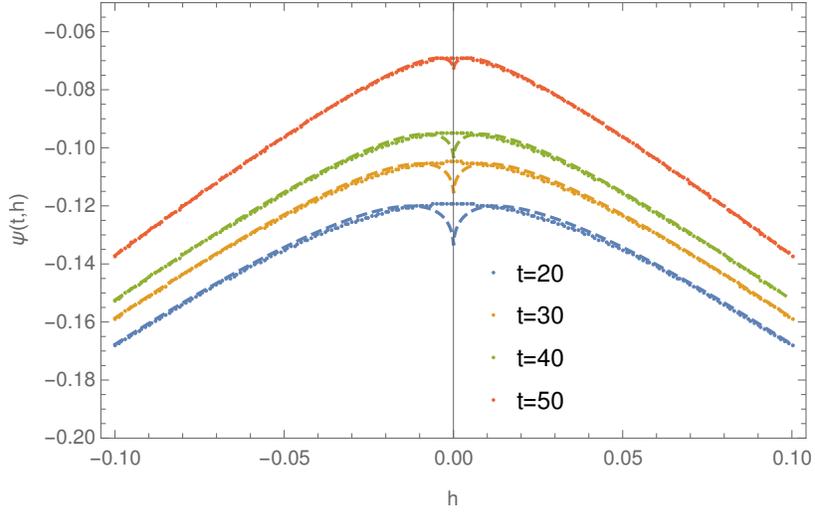

FIG. S6. The function $\psi(t,h)$ obtained via the QTM method (Eqs. (S68)) as a function of $h$ for different values of $t$. The dashed lines correspond to the expansion in (S64). Such an expansion does not capture the quadratic behavior at $h = O(t^{-1/2})$ [8].

trace in (S67) is dominated by the largest eigenvalue of the matrix $\tilde{F}_1 \tilde{F}_2$. Moreover, integrability is preserved by the space-time rotation, as can be checked explicitly from the Yang-Baxter relation (S2) (see also [7]); therefore, the dual transfer matrices can also be diagonalized by the method explained in Appendix C. Using the general expressions in (S13) and (S15), we arrive at the equation

$$t\tilde{p}(\lambda_a) + \sum_{b=1}^{t} \tilde{\Theta}(\lambda_a - \lambda_b) + \imath h = \pi(a - (t-1)/2), \qquad a = 0, \ldots, t-1 \tag{S68a}$$

$$\tilde{p}(\lambda) = \arctan(2\lambda) - \arctan(2\lambda/3), \quad \tilde{\Theta}(\lambda) = \arctan(\lambda) \tag{S68b}$$

$$\psi(t,h) = -\frac{1}{2} \sum_{a=0}^{t-1} \log\left(\frac{4\lambda_a^2 + 9}{16\lambda_a^2 + 4}\right) \tag{S68c}$$

The equations (S68a) can be solved efficiently with standard numerical tools for $t$ up to $10^4$; by inserting the solutions $\{\lambda_a\}_{a=0}^{t-1}$ in (S68c), one can efficiently obtain $\psi(t,h)$ for any values of $t$ and $h$. Finally, $\phi(t,s)$ is recovered by Legendre transform. A comparison between the exact $\psi(t,h)$ obtained with this method and the expansion (S64) is shown in Fig. S6.

## I. FREE MAGNON SCALING FUNCTION

In order to derive the scaling function $\kappa(x,s)$ given in Eq. (11) of the main text, we proceed as follows. First of all, since we are interested at the limit where both $t, L \to \infty$, we can neglect the trotterization involved in the matrix $\hat{M}$, and replace

$$K(t) = |t| \operatorname*{Tr}_s \left[\hat{M}^t\right] \approx |t| \operatorname*{Tr}_s \left[e^{-t\hat{H}}\right], \tag{S69}$$

where $\hat{H}$ is the XXX Hamiltonian. The scalng function is then defined by

$$\kappa(x,s) \equiv \ln\left[K(t)/t\right] \approx \ln \operatorname*{Tr}_s \left[e^{-t\hat{H}}\right], \tag{S70}$$

with $x = t/L^2$. This is essentially the free energy of the XXX model at 'temperature' $1/t$.

The single-particle excitations of this effective $\hat{H}$ are magnons, which behave effectively as bosons with quadratic dispersion relation $\varepsilon_k = Dk^2$ at small $k$. In a system of length $L$ with periodic boundary conditions the momenta are quantized according to $k = 2\pi n/L$ for integer $n$. At large $t$, the dominant contribution to $\kappa$ comes configurations

with only small-$k$ magnons excited, and at low densities, we can ignore interactions between magnons. Starting from (S70),

$$\kappa(x,s) = \ln \text{Tr}\left[e^{-t\hat{H}}\right] = \ln \prod_{k\neq 0} \sum_{n_k=0}^{\infty} e^{-t\varepsilon_k n_k} \tag{S71}$$

$$= \sum_{k\neq 0} \ln \frac{1}{1-e^{-tDk^2}} = -\sum_{n\neq 0} \ln\left[1-e^{-(2\pi n)^2 Dx}\right], \tag{S72}$$

where we ignore the $k=0$ ($n=0$) term as this corresponds to changes in the magnetization sector.

To reproduce the results of the main text, we note that the effective evolution $\hat{M}$ has dispersion $\varepsilon(\lambda) = -2\ln\cos k(\lambda)$; to leading order, one has $\varepsilon(k) = k^2$, i.e. $D=1$. The large $x$ limit of (S72) is dominated by the $n=1$ term; expanding the logarithm reproduces the limit $\kappa(x) \sim e^{-4\pi^2 Dx}$ for $x \gg 1$. Small $x$ corresponds to large $L$, as we are primarily interested in large $t$. In this case, we take the continuum limit, and the sum over momenta is replaced by an integral, which for either a $\cos k$ or $k^2$ dispersion gives $\kappa(x) \propto 1/\sqrt{x}$ for $x \ll 1$. Thus, the proposed scaling form (S72) reproduces the two limits $x \ll 1$ and $x \gg 1$ recovered in the main text with *exactly* the same coefficients to leading order.

In comparing with the numerical data at finite $q$, we leave the diffusion constant $D$ as a free parameter. Then, Eq. (S70) is also supported by its agreement with numerical study of $K(t)$.

---

[1] A. Chan, A. De Luca, and J. T. Chalker, Phys. Rev. X **8**, 041019 (2018).
[2] L. Faddeev, arXiv preprint hep-th/9605187 (1996).
[3] M. Gaudin, *The Bethe Wavefunction* (Cambridge University Press, 2014).
[4] M. Takahashi, *Thermodynamics of one-dimensional solvable models* (Cambridge University Press, 2005).
[5] C.-N. Yang and C. P. Yang, Journal of Mathematical Physics **10**, 1115 (1969).
[6] M. Suzuki, Physica A: Statistical Mechanics and its Applications **321**, 334 (2003).
[7] A. Klmper, Lecture Notes in Physics , 349379 (2004).
[8] M. Takahashi, Progress of Theoretical Physics Supplement **87**, 233 (1986)



\end{document}